\documentclass[useAMS,usenatbib,usegraphicx]{mn2e}

\usepackage{amssymb}
\usepackage{fixltx2e}

\newcommand{\cov}{\mathop{\rm Cov}\nolimits}
\newcommand{\sign}{\mathop{\rm sign}\nolimits}
\newcommand{\expect}{\mathbb{E}}
\newcommand{\disp}{\mathbb{D}}
\newcommand{\const}{\mathop{\rm const}\nolimits}

\newcommand{\FAP}{{\rm FAP}}
\title[Doppler noise in 55~Cancri]{Enhanced models for stellar Doppler noise reveal hints of
a 13-year activity cycle of 55~Cancri}

\author[R.V.~Baluev]{Roman V. Baluev\thanks{E-mail: r.baluev@spbu.ru}\\
Central Astronomical Observatory at Pulkovo of Russian Academy of Sciences, Pulkovskoje
shosse 65, St Petersburg 196140, Russia\\
Sobolev Astronomical Institute, St Petersburg State University, Universitetskij prospekt
28, Petrodvorets, St Petersburg 198504, Russia}

\begin{document}

\date{Accepted 2014 October 13.
      Received 2014 September 30;
      in original form 2014 May 15}

\pagerange{\pageref{firstpage}--\pageref{lastpage}} \pubyear{2014}

\maketitle

\label{firstpage}

\begin{abstract}
We consider the impact of Doppler noise models on the statistical robustness
of the exoplanetary radial-velocity fits. We show that the traditional model of the Doppler
noise with an additive jitter can generate large non-linearity effects, decreasing the
reliability of the fit, especially in the cases when a correleated Doppler noise
is involved. We introduce a regularization of the additive noise model that can gracefully
eliminate its singularities together with the associated non-linearity effects.

We apply this approach to Doppler time-series data of several exoplanetary systems. It
demonstrates that our new regularized noise model yields orbital fits that have either
increased or at least the same statistical robustness, in comparison with the simple
additive jitter. Various statistical uncertainties in the parametric estimations are often
reduced, while planet detection significance is often increased.

Concerning the 55~Cnc five-planet system, we show that its Doppler data contain significant
correlated (``red'') noise. Its correlation timescale is in the range from days to months,
and its magnitude is much larger than the effect of the planetary $N$-body perturbations in
the radial velocity (these perturbations thus appear undetectable). Characteristics of the
red noise depend on the spectrograph/observatory, and also show a cyclic time variation in
phase with the public Ca II H\&K and photometry measurements. We interpret this modulation
as a hint of the long-term activity cycle of 55~Cnc, similar to the Solar 11-year cycle. We
estimate the 55~Cnc activity period by $12.6\pm^{2.5}_{1.0}$~yrs, with the nearest minimum
presumably expected in 2014 or 2015.
\end{abstract}


\section{Introduction}
After the Doppler discovery of the famous planet 51~Pegasi~\emph{b} in
1995 \citep{MayorQueloz95}, a large progress was made in the exoplanets detection field,
and the number of robustly detected exoplanet candidates have recently exceeded $1000$
\citep{Schneider}. However, the development in this field have encountered a natural
barrier that is difficult to overcome. This barrier is related
to the stellar activity: even if we decrease the instrumental errors of the radial-velocity
(hereafter RV) measurements to a level of $1$~cm/s, we would not usually obtain the r.m.s.
below $1$~m/s due to the intrinsic astrophysical instabiltiy of even quiet
Solar-type stars. Only the quitest stars demonstrate the Doppler stability at a level below
$1$~m/s.

Any random noise can be suppressed, to a certain degree, by accumulating a large number of
observations. However, this path is increasingly hard, because the data analysis get
affected by an increasing number of various subtle noise effects that must
be therefore taken into account. In modern works on the analysis of exoplanetary Doppler
data we can see that the authors are in a fierce fight with each such systematic effect,
hoping to reduce the statistical uncertainties and detection thresholds, and
to eradicate various noise misinterpretation
errors \citep[e.g.][]{Dumusque12,Baluev13a,Hatzes13,Nelson14,Tuomi14}.

One of such effects is the correlated Doppler noise discovered recently in the precision RV
data for some stars, having a correlation timescale of $\sim 10$~d in average
\citep{Baluev11,Tuomi12,Baluev13a,FerozHobson14,Tuomi14}. This correlated, or `red',
RV noise have not got a good theoretic discussion yet, and depending on the case it may be
inspired by various instrumental effects as well as by stellar
activity phenomena (e.g. star spots).

In addition to the detection of previously unknown noise components, other
issues related to the noise models appear. These issues are related to the constructions of
robust models for the known noise components. Unfortanutely, so far we have no
good physical understanding of the structure of the Doppler noise at the accuracy level of
$\sim 1$~m/s. Models that are currently used to approximate the white or red RV noise often
represent only rather primitive formulae that were chosen mainly on the basis of
mathematical simplicity and convinience with the use of only minimal or no knowledge of the
underlying physics. Although some sources of the RV jitter like the stellar astroseismology
oscillations were identified and investigated long ago \citep[e.g.][]{OToole08}, the topic
of the RV jitter structure still represents a rather speculative and suggestive matter with
insuffiecient experimental/observational base, so the difficulties arising with accurate
modelling of the RV jitter are objectively defined.

Acknowledging this, we consider the issue of \emph{mathematical} usefulness of the RV noise
models. Definitely there are multiple models that can approximate the required general
physical behaviour equally well, but have considerably different
mathematical properties. Which one should be selected in practice? In the case
of exoplanetary Doppler data we should take a special care of the statistical robustness of
the derived RV fits. Under the fit robustness we understand here the degree of
non-linearity of the fitting task. Clearly, if we have two models that both adequately
reproduce the desired physical behaviour, we should select the one that generates a smaller
non-linearity effects, because unjustified non-linearity in the fitted model usually
only triggers various troubles.

What undesired consequences a non-linear model can impose? First of all, it is a violation
of the well-known asymptotic (large-sample) properties of the classic maximum-likelihood
estimation methods: the parametric estimations become considerably non-gaussian (sometimes
even multimodal with comparable modes), the likelihood-ratio test and all statistical tools
that rely on it become not applicable, and the maximum-likelihood estimations become biased
and loose their optimality (the asymptotic property of maximum efficiency). Also, highly
non-linear models require more iterations of the fitting algorithm, in comparison with the
well-linearizable cases. This results in an increased demand of computational resources.

One can object that modern statistical data-analysis methods like the Bayesian inference
\citep{Ford06,Gregory07a,Gregory07b,Tuomi13,Nelson14} can gracefully handle even
highly non-linear models, contrary to the classic maximum-likelihood approach. While
recognizing that the Bayesian methods indeed are currently the
best-developed statistical tools for solving non-linear estimation tasks, we emphasize the
practical difference between \emph{correct accounting for non-linearity}, which is
what Bayesian methods can do, and \emph{suppressing the non-linearity}
on the other hand. The ability of adequate handling of non-linear models does not mean that
the non-linearity effects are removed. Thus, even with the Bayesian approach we should try
to use, whenever possible, a well-linearizable model as a preferred alternative to a highly
non-linear one.

The construction of a well-linearizable and simultaneously physically adequate model
of Doppler noise is thus the primary goal of the paper. In Sect.~\ref{sec_issue} we explain
how a noise model can generate any non-linearity of the fit, how the traditional model with
an additive RV jitter may become highly non-linear, and what
is a well-linearizable noise model (only considering the white noise case). In
Sect.~\ref{sec_jitter} we describe the new ``regularized'' model of the white Doppler noise
that eliminates the weaknesses of the additive model. In Sect.~\ref{sec_rednoise} we
consider the regularity issues raised by the red noise models. In Sect.~\ref{sec_prdg}
we discuss several issues concerning various periodograms, and how they are affected
by non-linearity. We also give there slightly extended results
concerning the periodogram significance levels for models involving the red noise.
In Sect.~\ref{sec_simul} we present testcase Monte Carlo simulations demonstrating that the
regularized noise model may lead to a dramatic improvement of the fit robustness in some
practical cases. In Sect.~\ref{sec_55Cnc} we present the results of the new analysis of the
Doppler data for the star 55~Cancri that hosts a five-planet system (using the new noise
modelling technique). In Sect.~\ref{sec_activity} we discuss the hints of a decade activity
cycle revealed in these Doppler data and the relationship of these results
with the observations from other long-term monitoring projects that involved 55~Cnc.

\section{White noise models and non-linearity}
\label{sec_issue}
Let us assume that we deal with the time series containing $N$
measurements $x_i$ (e.g. radial velocities), acquired at the times $t_i$, and having the
random errors $\epsilon_i$. The model that $x_i$ should be fitted with is designated
as $\mu(\btheta,t)$. We now consider several models of the noise
$\epsilon_i$ and the consequences they can carry for the fit of $\mu$.

In this section we assume that $\epsilon_i$ represent white Gaussian noise, i.e. that they
are independent normally distributed variates. In this case to set a `model'
of the noise means to express the variances $\disp\epsilon_i$ through some noise parameters
$\bmath p$ by means of a relationship
\begin{equation}
\disp\epsilon_i =\sigma_i^2(\bmath p),
\end{equation}
where the function $\sigma_i^2(\bmath p)$ represents the noise model that we are
speaking of. This framework was already considered in general terms, e.g. by
\citep{Baluev08b}, where it was suggested to fit the curve
parameters $\btheta$ and the noise parameters $\bmath p$ simultaneously by means
of the maximum-likelihood approach. This can be achieved through the maximization of the
log-likelihood function:
\begin{eqnarray}
\ln\mathcal L = -\frac{1}{2} \sum_{i=1}^N \left[ \ln \sigma_i^2(\bmath p) +
 \frac{r_i^2(\btheta)}{\sigma_i^2(\bmath p)} \right], \quad r_i=x_i-\mu(\btheta,t_i), \nonumber\\
\{\btheta^*,\bmath p^*\} = \arg \max \ln\mathcal L(\btheta,\bmath p).
\label{loglik}
\end{eqnarray}

Here we consider a few special cases of the noise model $\sigma_i^2(p)$, depending
on a single scalar parameter $p$.

The classic models of the noise assume that $\epsilon_i$ either have known
standard deviations $\sigma_i$ or at least known statistical weights $w_i\propto
1/\sigma_i^2$. In the first case we have
\begin{eqnarray}
\sigma_i^2 = 1/w_i = \const, \label{fn} \\
\ln\mathcal L = \frac{1}{2} \left[- \chi^2(\btheta) + \sum_{i=1}^N\ln w_i\right], \label{loglik-fn} \\
\chi^2(\btheta) = \sum_{i=1}^N w_i r_i^2(\btheta) \label{chi2}.
\end{eqnarray}
Now the task is reduced to the classic least-square fitting as
$\btheta^* = \arg\min\chi^2(\btheta)$.

In the second case we deal with the arbitrarily scalable noise:
\begin{eqnarray}
\sigma_i^2(\kappa) = \kappa/w_i,
\label{mult} \\
\ln\mathcal L = \frac{1}{2}\left[- N\ln\kappa - \frac{\chi^2(\btheta)}{\kappa} + \sum_{i=1}^N\ln w_i \right],
\label{loglik-mult}
\end{eqnarray}
where $\kappa$ is an unknown parameter and $\chi^2$ is given by~(\ref{chi2}). We also call
this classic noise model as `multiplicative' one. Clearly, now the likelihood function can
be maximized over $\btheta$ and $\kappa$ separately, so that we have $\kappa^* =
\chi^2(\btheta^*)/N$, and $\btheta^*$ is again
obtained through the least-squares fitting.\footnote{The reader may notice that this
estimation of $\kappa$ is actually biased, while the classic \emph{unbiased}
expression should be $\kappa=\chi^2/(N-m)$, where $m$ is the number of the curve
parameters, $m=\dim\btheta$. The necessary bias correction is performed in
\citep{Baluev08b} by modifying the likelihood function~(\ref{loglik}), but we do not
highligt this change here, keeping more simple formulae for demonstration.}

In this paper we are mainly concerned by the effects of the non-linearity of the task. A
strictly linear curve model $\mu(\btheta,t)=\btheta\cdot\bvarphi(t)$, where $\bvarphi$
represent some known functional base, would imply that $\chi^2(\btheta)$
is expressed though a quadratic form, so that its minimization becomes trivial. If the
model $\mu$ is formally non-linear, but the function $\chi^2(\btheta)$ still can
be approximated by a positive-definite quadratic form, we say that the least-square task is
well-linearizable. If $\chi^2(\btheta)$ cannot be approximated
by a quadratic function, especially if it possesses multiple local minima, we say that the
associated least-square task is essentially non-linear. However, it is not trivial
to classify the noise models $\sigma_i^2(p)$ in the same manner. Here we should base on the
likelihood function~(\ref{loglik}) rather than on the noise model itself. Namely, we should
say that the noise model is `linear' if the log-likelihood $\ln\mathcal L$ is expressed by
a quadratic form in terms of the noise parameter $p$. The model $\sigma_i^2(p)$ is not
supposed to be linear here in the strict mathematical sense.

Now let us return to the classic multiplicative model~(\ref{mult}). Is it linear in this
sense or not? Apparently, even if we consider $\btheta$ fixed, the
relevant log-likelihood~(\ref{loglik-mult}) is not quadratic in $\kappa$. However,
we may note that this log-likelihood always has only a single maximum in $\kappa$, and
it tends to $-\infty$ both for $\kappa\to +0$ and $\kappa\to +\infty$.
Therefore, for a fixed $\btheta$ we can easily find a parametric replace
$\kappa\to p$, such that the log-likelihood function is quadratic in $p$. We cannot yet say
that the inferred likelihood function is strictly quadratic in both $\btheta$ and $p$,
because the replace $\kappa\to p$ also depends on $\btheta$. However, in practice we may
consider the values of $\btheta$ in a small ($\sim 1/\sqrt N$) vicinity around the
best fitting solution, so we may say that~(\ref{loglik-mult}) can be
rather accurately, although not exactly, represented by some quadratic function in
$p(\kappa)$. Thus, the multiplicative noise model~(\ref{mult}) can generate little
non-linearity effects. The cumulative non-linearity of the maximum-likelihood task is thus
dominated by the non-linearity coming from the curve model $\mu$.

After that, let us consider the following `additive' noise model from \citep{Baluev08b}:
\begin{eqnarray}
\sigma_i^2(p) = \max(p + \sigma_{i,\rm meas}^2,0), \label{add} \\
\ln\mathcal L = -\frac{1}{2} \sum_{i=1}^N \left[ \ln (p + \sigma_{i,\rm meas}^2) +
 \frac{r_i^2(\btheta)}{p + \sigma_{i,\rm meas}^2} \right], \label{loglik-add}
\end{eqnarray}
where $\sigma_{i,\rm meas}$ designate some known uncertainties, while the unknown parameter
$p$ is responsible for the additional `jitter'. The model~(\ref{add}) was originally
constructed assuming that $p$ physically represents the variance of some
additional noise component, implying $p>0$ \citep{Wright05}. However, here we do not
require $p$ to be necessarily positive. The model~(\ref{add}) may remain quite
meaningful even for $p<0$. The negative value of $p$ means that the real
errors $\epsilon_i$ are systematically smaller than the stated uncertainties $\sigma_{i,\rm
meas}$, because e.g. these $\sigma_{i,\rm meas}$ were overestimated due to
some imperfections of the spectrum analysis pipeline. In practice such cases exist
indeed \citep{Baluev08b}, and therefore the ability of the noise model~(\ref{add})
to handle such cases is pretty desirable, even though its original
motivation becomes non-physical for $p<0$.

\begin{figure}
\includegraphics[width=84mm]{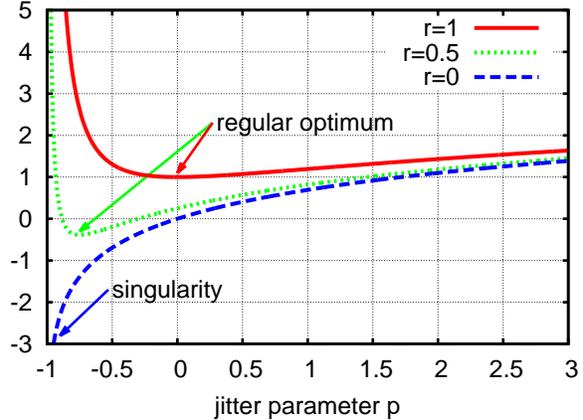}
\caption{Graphs of the elementary term in the likelihood function~(\ref{loglik}) for the
additive noise model. The plots correspond to the three labelled values of the residual $r$
and to $\sigma_{i,\rm meas}=1$.}
\label{fig_sing}
\end{figure}

Nonetheless, not every negative value of $p$ is meaningful. We may notice an undesired
irregularity in the likelihood function near $p=-\min \sigma_{i,\rm meas}^2$, which is the
minimum admissible value of $p$. When $p$ approaches to this value, one of $\sigma_i^2(p)$
tends to zero, but contrary to the multiplicative model, the others $\sigma_i^2(p)$ remain
separated from zero. This implies that the relative weight of the single observation
with the minumum $\sigma_{i,\rm meas}$ increases infinitely, so this observation plays
an increasingly dominant role. This is obviously an undesired behaviour: we have
no any reasonable justification to infinitely increase the contrast between the weights.
This singularity does not always show itself up. Normally, the residual $r_i$ corresponding
to the observation with minimum $\sigma_{i,\rm meas}$ does not vanish, and the values of
$p$ near this limit are naturally penalized by the
term $r_i^2/\sigma_i^2(p)$ in~(\ref{loglik}). This term tends to $-\infty$
at a greater speed than the other term, $\log\sigma_i^2(p)$, tends to $+\infty$. The
resulting log-likelihood thus tends to $-\infty$, indicating an unsuitable solution.
However, we obtain a degeneracy when this residul $r_i$ vanish, leaving only the
logarithmic term in the game. In this case the entire log-likelihood function would tend to
$+\infty$, indicating an absolutely (!) preferrable solution.

This singularity effect is illustrated in Fig.~\ref{fig_sing}. As we can
see, the singularity is very sensitive with respect to the residual $r_i$, so to
be actually trapped, the likelihood maximization algorithm should meet in its path a small
enough residual $r_i$ simultaneously with a negative value of $p$. In fact this singularity
effect was already mentioned in \citep{Baluev08b}, where also a workaround solution to the
issue was suggested. The workaround was to cut off the singularities by truncating the
additive model~(\ref{add}), replacing it with
\begin{eqnarray}
\sigma_i^2(p) = \left\{\begin{array}{l}
                        p + \sigma_{i,\rm meas}^2, \quad p>=(a^2-1)\sigma_{i,\rm meas}^2 \\
                        0, \quad p<(a^2-1)\sigma_{i,\rm meas}^2.
                       \end{array}
                \right.\nonumber\\
 a<1
\label{addtrunc}
\end{eqnarray}
and using the general likelihood~(\ref{loglik}). This forbids the maximum-likelihood fitter
to travel to the domain where any of $\sigma_i(p)$ becomes smaller
than $a\sigma_{i,\rm meas}$. For example, the truncation with $a=0.1$ is applied internally
in PlanetPack \citep{Baluev13c}, i.e. it is not allowed to reduce the uncertainties more
than by the factor of $10$. In practice this pretty mild condition allows the fitter to be
never trapped in the singularity, although it is still possible to have the best fitting
value of $p$ lying at the very boundary of the truncation.

Although the truncated model~(\ref{addtrunc}) solves the numerical issues of the fitting by
cutting off the likelihood singularites themselves, it does not eliminate the severe
non-linearity effects that expand over a wider domain around the
singularity. The observations with smaller values of $\sigma_{i,\rm meas}$ still receive a
disproportionally increasing relative weights for $p<0$ without good physical reasoning.
The other observations cannot have significant effect on the fit in such a case.
This basically means that the effective number of observations involved in the game gets
decreased, pumping up the non-linearity of the task. However, we do not want to just ban
the negative values $p$ completely, because as we disscussed above, negative $p$ allows to
handle practical cases with overestimated errors. Besides, any abrupt cutoffs of the
parameters introduce other non-linearity effects, so the model~(\ref{addtrunc}) is still
just an ugly workaround rather than a neat solution to the issue.

To suppress the non-linearity effects generated by the additive noise model for $p<0$ we
should construnct a new `regularized' noise model with better mathematical properties in
this domain. In the domain $p>0$ this new model should behave like the
additive model~(\ref{add}), while for $p<0$ it should behave similarly
to the multiplicative model~(\ref{mult}), keeping the weight ratios of the
observations constant.

The graph of the additive model~(\ref{add}) consists of two straight rays:
$\sigma_i^2(p)=p+\sigma_{i,\rm meas}^2$ for $p>-\sigma_{i,\rm meas}^2$ and
$\sigma_i^2(p)=0$ for $p<-\sigma_{i,\rm meas}^2$. Given this, we
define our regularized model as a hyperbola constructed on these rays as
on asymptotas. Mathematically, such regularized model may be expressed as
\begin{eqnarray}
\sigma_i^2(\kappa) = \sigma^2_{\rm scale} \psi_{\rm reg}(\kappa,\tau_i), \qquad
\tau_i=\frac{\sigma_{{\rm meas},i}}{\sigma_{\rm scale}}, \nonumber\\
\psi_{\rm reg}(\kappa,\tau_i) = \frac{1}{2}\left( \kappa+\tau_i^2 + \sqrt{(\kappa+\tau_i^2)^2+(2a\tau_i)^2} \right),
\label{wnreg}
\end{eqnarray}
with the likelihood given by~(\ref{loglik}). Here $\sigma_{\rm scale}$ is some dimensional
scale factor, $\kappa$ is a fittable adimensional jitter parameter, and $a$ is a small
regularization parameter. For $a=0$ the model~(\ref{wnreg}) is strictly equivalent
to the additive jitter model. For non-zero $a$ and $\kappa\to +\infty$ the
model~(\ref{wnreg}) is asymptotically equivalent to the additive model, with the
relative error decreasing as $\sim 1/\kappa^2$. For $\kappa\to -\infty$ we
obtain $\sigma_i^2(\kappa)
\sim (a\sigma_{i,\rm meas})^2/|\kappa|$, which becomes equivalent to the
multiplicative model. Also, for~(\ref{wnreg}) we
have $\sigma_i^2(-a^2)=\sigma_{\rm scale}\tau_i^2 = \sigma_{i,\rm meas}^2$, meaning that
our regularized model for $\kappa=-a^2$ coincides exactly with the additive model for $p=0$
(regardless of the values of $\tau_i$).

\section{Regularized model of the additive jitter}
\label{sec_jitter}
The regularization~(\ref{wnreg}) is not a unique choice, although this is one of the
simplest mathematical functions that suits our needs. Note that the original
multiplicative~(\ref{mult}) and additive~(\ref{add}) models can be represented in the same
form as in~(\ref{wnreg}), substituting the following functions in place of
$\psi_{\rm reg}$:
\begin{eqnarray}
\psi_{\rm mult}(\kappa,\tau_i) &=& \kappa \tau_i^2, \quad w_i = 1/\sigma_{i,\rm meas}^2 \nonumber\\
\psi_{\rm add}(\kappa,\tau_i) &=& \max(\kappa + \tau_i^2, 0), \quad p=\kappa\sigma_{\rm scale}^2.
\label{wnmultadd}
\end{eqnarray}
We can see that in comparison with $\psi_{\rm add}$, the function $\psi_{\rm reg}$ offers a
smooth non-breaking decrease of the variance for $\kappa\to -\infty$, which removes any
pecularities from the likelihood function and thus may reduce the effects of non-linearity.
Note that the relative contrast between the weights of the observations keeps bounded by
the contrast between $\sigma_{i,\rm meas}^2$. Thus we also eliminated the pathological
property of the additive model that forced only one or a few of the RV measurements to
dominantly contribute in the final fit if $\kappa<0$.

\begin{figure*}
\includegraphics[width=0.45\textwidth]{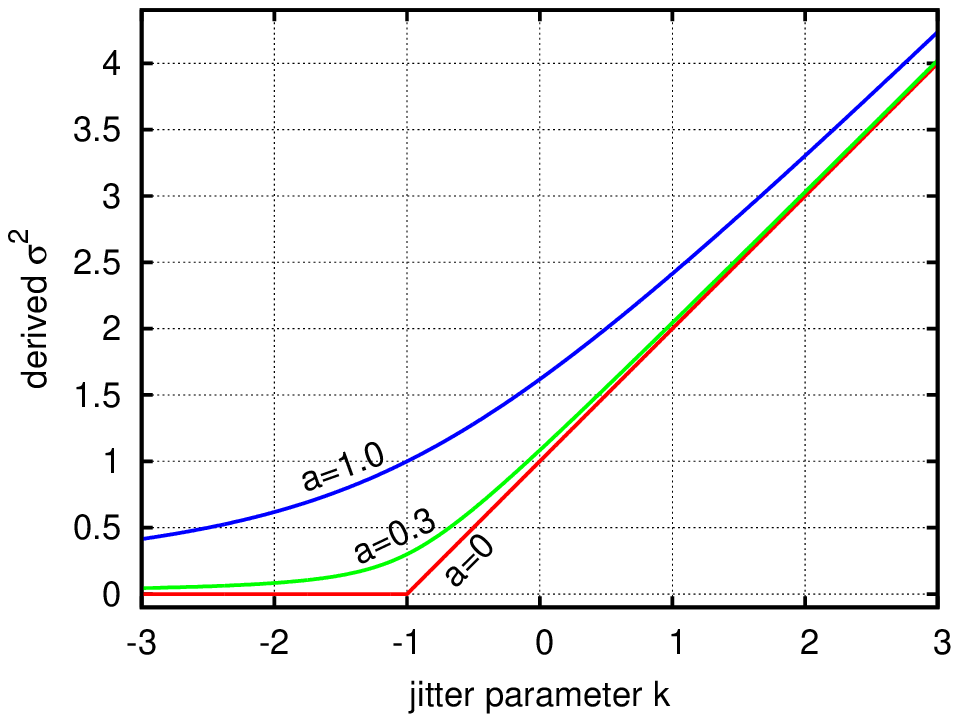}
\includegraphics[width=0.45\textwidth]{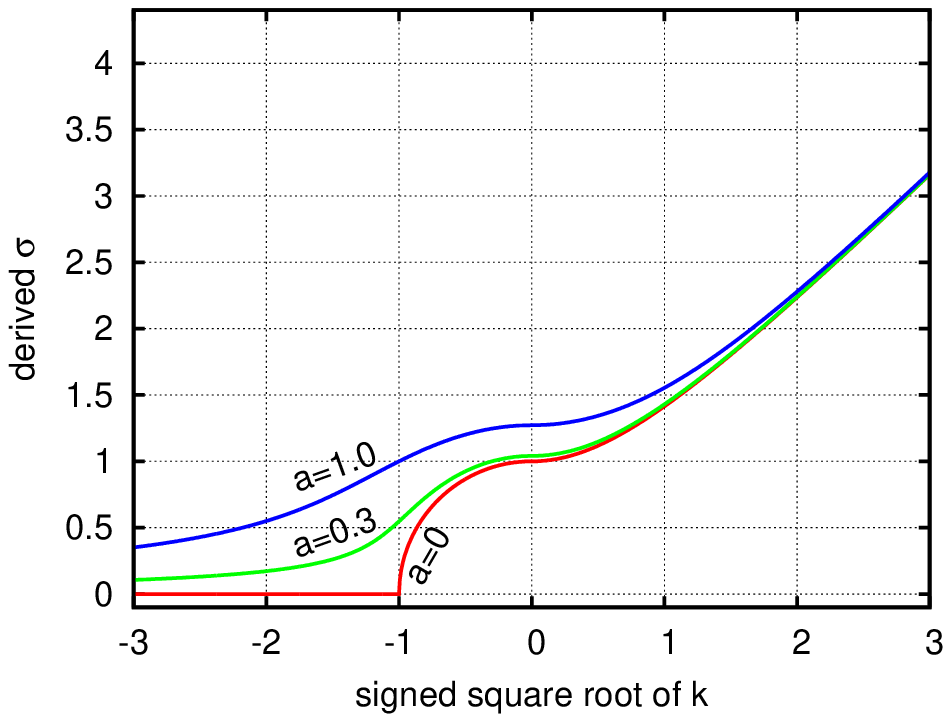}
\caption{Regularized jitter model for three marked values of the regularization parameter
$a$. The left frame represents the graph of the modelled total variance $\sigma^2$ as a
function of the jitter parameter $\kappa$, and the right frame shows the implied standard
deviation $\sigma$ as a function of the signed square-root $\sqrt{|\kappa|}\, \mathop{\rm sign}
\kappa$. Case $a=0$ corresponds to the additive noise model. It is assumed here that $\tau=1$
and $\sigma_{\rm scale}=1$ (see text for the detailed discussion).}
\label{fig_regjit}
\end{figure*}

The regularized model~(\ref{wnreg}) has two fine-control parameters,
$\sigma_{\rm scale}$ and $a$. They are not independed: the replacement $(\sigma_{\rm
scale}, a, \kappa) \mapsto (c \sigma_{\rm scale}, a/c, \kappa/c^2)$ preserves the model.
Therefore, we may set $\sigma_{\rm scale}$ to any reasonable dimensional value and
then select the non-dimensional parameter $a$. We set $\sigma_{\rm scale}$ equal
to the geometric mean of $\sigma_{i,\rm meas}$, and then after some experimenting we
find that good compromise values for $a$ are located in the range from $a=0.3$ to $a=1.0$.
The graphs of the regularized model~(\ref{wnreg}) for a few values of $a$ are plotted in
Fig.~\ref{fig_regjit}.

The jitter parameter $\kappa$ in~(\ref{wnreg}) and~(\ref{wnmultadd}) is just a dummy
adimensional variable. For the multiplicative and additive jitter models, we
define the relevant \emph{dimensional} RV jitter using the formula
\begin{equation}
\sigma_\star = \sigma_{\rm scale} \sqrt{|\kappa|}\, \sign \kappa,
\label{jitdef}
\end{equation}
where $\kappa$ is the parameter of~(\ref{wnmultadd}). Here we follow the convention that
negative $\sigma_\star$ corresponds to negative $\kappa$ (implying overestimated
$\sigma_{i,\rm meas}$). The same definition could be
used for the regularized model~(\ref{wnreg}) too. However, in this case we may want to make
the regularized and the additive jitter models interchangeable in the sense that switching
between them should not change the implied jitter estimation too much. With~(\ref{jitdef})
this requirement is not satisfied. For example, as we noted above, the regularized model
for $\kappa=-a^2$ coincides with the additive one for $p=0$. In this case we would
expect to have $\sigma_\star=0$, but we obtain from~(\ref{jitdef}) a
non-zero value $\sigma_\star=-a\sigma_{\rm scale}$, which is not
always satisfactory. For smaller negative $\kappa$, the bias in $\sigma_\star$
is further increased, e.g. the case $\kappa\to-\infty$ should correspond
to the effective jitter of about $-\sigma_{\rm scale}$, rather than to minus
infinity implied by~(\ref{jitdef}).

Let us consider this effect in more details. From~(\ref{wnreg}) it follows that
\begin{equation}
\kappa = \frac{\sigma_i^2}{\sigma_{\rm scale}^2} - \tau_i^2 - \frac{a^2\tau_i^2\sigma_{\rm scale}^2}{\sigma_i^2}.
\label{wnreg-inv}
\end{equation}
Now let us assume that $\sigma_i^2$ actually obey the additive model~(\ref{wnmultadd}) for
some $\kappa=\hat \kappa$, and we try to approximate these $\sigma_i^2$ using a slightly
different model~(\ref{wnreg}). That is, we seek a replacement $\kappa(\hat\kappa)$
such that $\psi_{\rm reg}(\kappa)\approx \psi_{\rm
add}(\hat\kappa)$. From~(\ref{wnreg-inv}) follows that for each $\sigma_i$
\emph{individually} such a representation can be made exact if we substitute
in~(\ref{wnreg})
\begin{equation}
\kappa = \hat \kappa - \frac{a^2\tau_i^2}{\hat \kappa + \tau_i^2}.
\label{par_bias}
\end{equation}
However, we should find a single \emph{common} value of $\kappa$ approximating all
$\sigma_i$ simultaneously. Such value of $\kappa$ can be expressed using the
same formula~(\ref{par_bias}), but replacing all $\tau_i$ by some their mean value
$\tau_*$. Moreover, if we choose $\sigma_{\rm scale}$ wisely, this mean value $\tau_*$
should be close to unit. Therefore, replacing $\tau_i$ in~(\ref{par_bias}) by unit, we
obtain a suitable `best fitting' representation for $\kappa$:
\begin{equation}
\kappa = \hat \kappa - \frac{a^2}{\hat \kappa + 1}.
\label{par_bias0}
\end{equation}

Thus the entire procedure of determining the jitter estimation looks as follows. First
of all, we perform the maximum-likelihood fit of the regularized model~(\ref{wnreg}) via
the parameter $\kappa$. Then we derive from~(\ref{par_bias0}) the matching value
of $\hat\kappa$:
\begin{equation}
\hat\kappa = \psi_{\rm reg}(\kappa,1)-1,
\label{par_ibias0}
\end{equation}
which represents the argument of an additive model that approximates the fitted regularized
one. Finally, we substitute this value $\hat\kappa$ in place of $\kappa$
into the definition~(\ref{jitdef}). When $\sigma_\star$ is defined using this procedure,
its global minimum value is $-\sigma_{\rm scale}$, which maps to $\kappa=-\infty$. The
value $\kappa=-a^2$ maps to $\hat\kappa=0$, and this now correctly corresponds
to $\sigma_\star=0$. For large positive $\kappa$ this corrective offset becomes negligible.

\begin{figure*}
\includegraphics[width=0.45\textwidth]{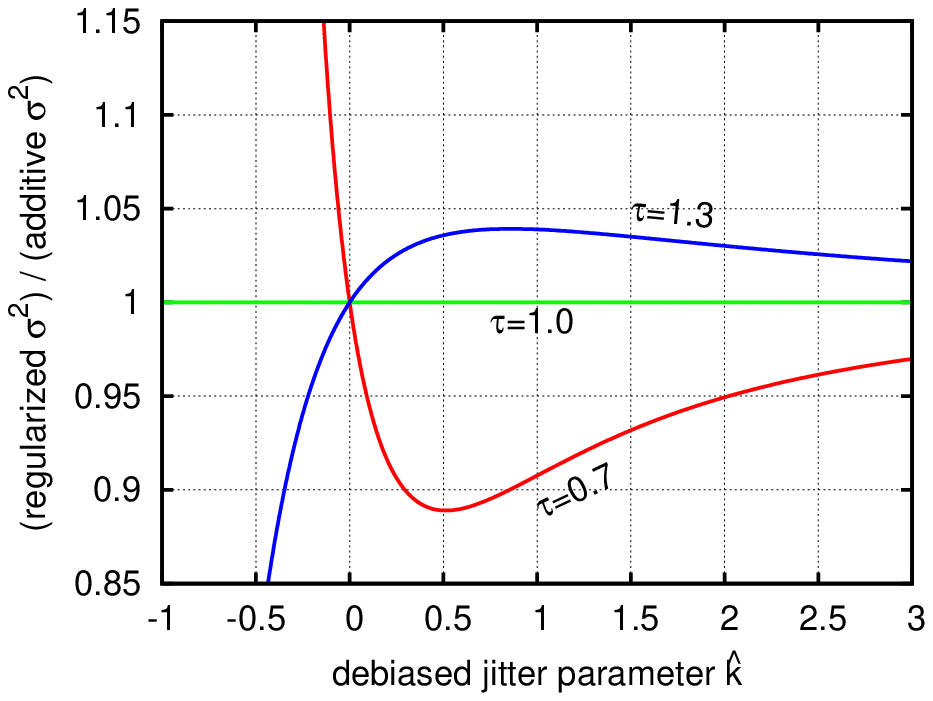}
\includegraphics[width=0.45\textwidth]{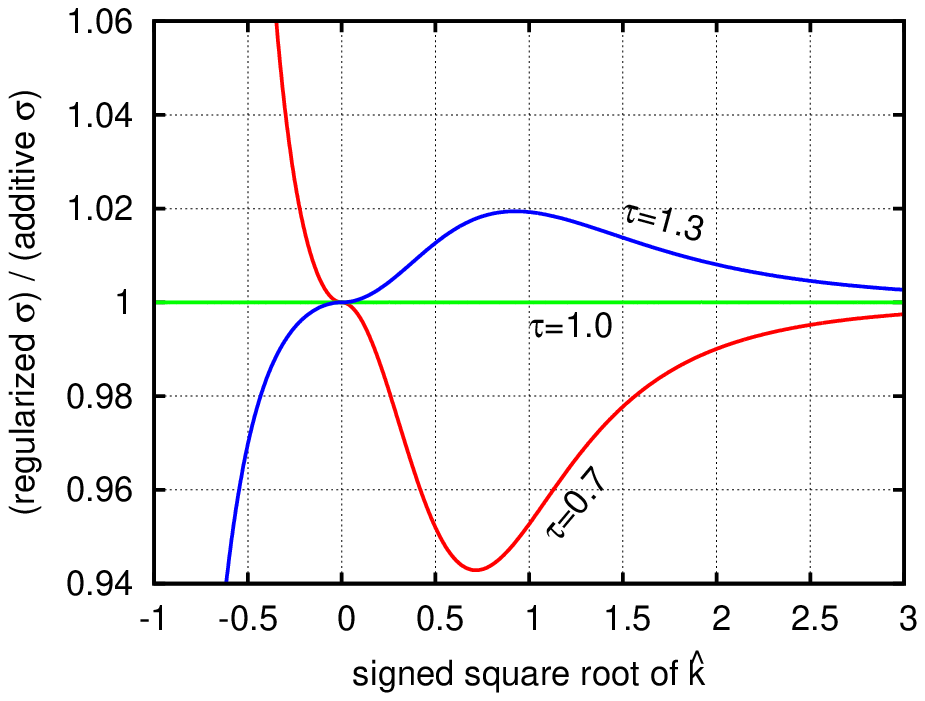}
\caption{Error of the regularized jitter model, remaining after compensating the average
bias of the jitter parameter $\kappa$. The left frame shows the ratio $\psi_{\rm
reg}(\kappa(\hat\kappa))/\psi_{\rm add}(\hat\kappa)$ as a function of the debiased
parameter $\hat\kappa$ (see text for details), and the right frame shows the implied ratio
of standard deviations as a function of the signed square-root
$\sqrt{|\hat\kappa|}\, \mathop{\rm sign} \hat\kappa$. The curves are plotted for three
labelled values of the uncertainty parameter $\tau$. It is also assumed here that $a=1$ and
$\sigma_{\rm scale}=1$.}
\label{fig_regjit_bias}
\end{figure*}

This also implies that the actuall modelling errors of~(\ref{wnreg}) are smaller than it
may seem from Fig.~\ref{fig_regjit}, because in this figure no correction was applied to
the argument $\kappa$. Considering the regularized model in~(\ref{wnreg}) as a function of
$\hat\kappa$, like $\psi_{\rm reg}(\kappa(\hat\kappa))$, we would obtain
\emph{exact} coincidence with the additive noise model
$\psi_{\rm add}(\hat\kappa)$ for $\tau=1$. For $\tau=0.7$ or $\tau=1.3$, and with $a=1$, we
obtain that the modelling errors of $\sigma_i^2$ remain below $12\%$ for $\kappa>0$ (see
Fig.~\ref{fig_regjit_bias}). Therefore, even $a=1$ is still a small enough value
that allows~(\ref{wnreg}) to efficiently mimic the additive model in the domain $\kappa>0$.
However, for $\kappa<0$ the divergence grows quickly, allowing the regularized model
to have much better mathematical properties. These properties lead to an
improved robustness of various statistical estimations and to a better predictability of
various statistical thresholds.

\section{Regularity of the red noise models}
\label{sec_rednoise}
The activity-related Doppler variations in solar-like stars were investigated relatively
long ago \citep[e.g.][]{SaarDonahue97,Saar98}, but the available RV data amount and quality
disabled any detailed modelling of this Doppler noise until recent years. In the most of
the exoplanetary RV-analysis works in the past, the Doppler ``jitter'' was modelled as a
white noise. Now it is clear that a well detectable correlated (or ``red'') noise is often
present in the exoplanetary Doppler data. The number of known stars with significant
red-noise component in the RV data
is growing \citep{Baluev11,Baluev13a,Tuomi12,FerozHobson14,Tuomi14}, and these
examples demonstrate the typical noise correlation timescales of $\sim 10$~d in average.
Apparently, this red noise is excitated by the stellar
activity effects \citep{Dumusque12,Hatzes13,Robertson14}.

In \citep{Baluev11,Baluev13a} and \citep{FerozHobson14} the red noise in Doppler
data was taken into account by means of adopting a realistic (although
rather arbitrarily selected) parametrized model for its covariance structure. The following
two-component model of the correlated RV noise was applied:
\begin{eqnarray}
V_{ij} = \cov(x_i,x_j) = \sigma_{i,\rm wht}^2(p_{\rm wht})\delta_{ij} + V_{ij,\rm red}(p_{\rm red},\tau), \nonumber\\
V_{ij,\rm red} = p_{\rm red} R_{ij}(\tau), \quad
R_{ij} = \rho\left(\frac{t_i-t_j}{\tau}\right), \nonumber\\
\sigma_{i,\rm wht}^2 = p_{\rm wht} + \sigma_{i,\rm meas}^2,
\label{rednoise}
\end{eqnarray}
where $p_{\rm wht}$, $p_{\rm red}$, and $\tau$ are free fittable parameters, while
$\rho(t)$ is the adopted shape of the correlation function. We
used $\rho(t)=\exp(-|t|)$, corresponding to the so-called ``red'' noise.

The likelihood of the correlated data is different
from~(\ref{loglik}). Approximating the red noise by a stationary Gaussian random process
with the covariance matrix~(\ref{rednoise}), the likelihood can be expressed by
\begin{equation}
\ln\mathcal L = -\frac{1}{2}\left(\ln\det\mathbfss V + \bmath r^{\rm T} \mathbfss V^{-1} \bmath r \right),
\label{loglik-corr}
\end{equation}
with the residuals vector $\bmath r$ same as in~(\ref{loglik}). This likelihood is a
function of three noise parameters $p_{\rm wht}$, $p_{\rm red}$, and $\tau$, and
of the usual parameters of the RV curve, $\btheta$. The joint estimation
of all these parameters can be obtained by finding the maximum of~(\ref{loglik-corr}),
as usually.

We can see that the noise model~(\ref{rednoise}) is potentially vulnerable with respect to
the irregularities discussed above, because we did not put any constraints on $p_{\rm wht}$
or $p_{\rm red}$, except for an implicit condition that the resulting noise covariance
matrix $\mathbfss V$ should be positive definite. Note that with the model~(\ref{rednoise})
the cases of $p_{\rm wht}<0$ become very frequent, because a significant fraction of the
available residulas scatter may be consumed by the ``red'' term with $p_{\rm
red}$. Moreover, because we now have two noise components, we may face two types
of singularities. They appear in the domains $p_{\rm wht}<0$ or $p_{\rm red}<0$. A
singularity in the likelihood~(\ref{loglik-corr}) formally occures when $\det\mathbfss
V$ passes zero. The singularity of the white component can be removed by the truncated
model~(\ref{addtrunc}) or preferably by the new regularized model~(\ref{wnreg}).

What concerns the red noise component, here the solution is even more simple: we
should just forbid $p_{\rm red}$ to attain negative values, as they are non-physical. In
this case we do not have any justification to possibly allow $p_{\rm red}<0$, like
we had with $p_{\rm wht}$. The constraint $p_{\rm red}\geq 0$ can be
easily implemented e.g. by replacing $p_{\rm red}=\sigma_{\star,\rm red}^2$ and treating
$\sigma_{\star,\rm red}$ as a primary fittable parameter.

\section{Likelihood-ratio periodograms}
\label{sec_prdg}
Periodograms represent an extremely important analysis tool for Doppler exoplanets
detections. Besides, periodograms are rather sensitive to the non-linearity of the adopted
models, see e.g. \citep{BaluevBeauge14}. In fact, we originally suspected the non-linearity
issues of the additive model~(\ref{add}) because of the too frequent
disagreement between the simulated periodogram significance levels and their
expected analytic approximations.

The general likelihood-ratio periodogram was formally defined in \citep{Baluev08b}, and it
basically represents the difference between maximum values of log-likelihood
obtained for two RV curve models, with and without a probe sinusoidal signal. This
is a general definition valid both for various noise models, including white or correlated.

We would like to highlight an important internal property of these periodograms that often
makes them superior over those periodograms that are still used in the majority of other
exoplanetary data-analysis works. This property becomes important when we deal
with multi-planetary fits, e.g. when we have already detected one planet, and need
to verify whether the data contain yet another signal. In this case we
should compare the base (or null) model, involving only the signal of the known planet, and
the alternative model involving the signals of the known planet + additional sinusoid. The
wide-spread solution to this issue is to fit the base model, compute the residuals of this
best fit, and then treat these residuals as a new standalone time series, applying to it
the same periodogram that was used to detect the first planet (i.e., assuming a zero
base model). It was clearly demonstrated by
\citep{Anglada-Escude12} that this approach is inoptimal and leads to a decreased detection
power. The second periodogram here does not take into account the possibility that the
first best fit may be inadequate or inaccurate due to the contribution of the putative
second planet. Thus, the residuals obtained after the first fit may be corrupted
by the inaccuracies of the base model. A better solution is to adaptively adjust the
parameters of the first planet in the process of computing the second periodogram. In other
words, the second periodgram should involve the fit of the \emph{original} data (instead of
the single-planet residuals) with the entire \emph{two-planet} model (i.e. first
planet + putative sinusoid, rather than just the sinusoid), comparing this fit with the fit
of only the base model. This modification leads us to the likelihood-ratio statistic and
the associated likelihood-ratio periodograms from \citep{Baluev08b}.

We call such periodograms as ``residual periodograms'', contrary
to the usual ``periodograms of residual'', while \citet{Anglada-Escude12} calls them
``recursive periodograms''. The value of the residual periodogram is always greater than or
equal to the value of the periodogram of the fixed
residuals, because the first one involves a complete maximization of the
likelihood function, while the second one corresponds to another, non-maximum, value of the
likelihood. Simultaneously, the noise level of the residual periodograms remains
approximately the same, making the residual periodograms more sensitive to faint
signals. This was practically demonstrated by \citet{Anglada-Escude12}.

Note that when the base model is expressed by just a constant the same issue leads
to the construction of the ``floating-mean periodogram''
\citep{FerrazMello81,Cumming99,ZechKur09}. The floating-mean periodogram is a more
efficient tool than the \citet{Lomb76}-\citet{Scargle82} periodogram of a centred time
series. Basically, the residual likelihood-ratio periodograms just
represents a further generalization of the floating-mean periodogram to the cases with more
complicated models of the RV curve and non-trivial models of the RV noise.

The likelihood-ratio periodograms also allow an extension of another type, useful when
dealing with heterogeneous data, e.g. RV data for the same star coming
from independent teams. In this case we may be interested in detecting
possible descrepancies between the different data sets. This can be achieved
by constructing periodograms referring -- in some sense -- to individual data sets. We
construct such dataset-specific periodograms by comparing the base model and the
alternative model, obtained by adding a sinusoidal signal to the base model \emph{only for
data points that belong to a particular subset}. For other data points the alternative
model is set equal to the base one. The resulting likelihood ratio statistic allows to
reveal variations belonging to only a particular dataset, still using the full statistical
power of the entire time series to fit the base model. This method was introduced in
\citep{Baluev11} and its practical efficiency was further demonstrated
in \citep{BaluevBeauge14}. Clearly, this approach is more efficient then
e.g. a plain restriction of the analysis to only a single RV data set.

The crucial matter in the periodogram analysis is the determination of the statistical
significance of the periodogram peaks that we reveal. The analytic
theory for the periodogram significance levels is given in \citep{Baluev08a}. Summarizing,
the false alarm probability $\FAP$ associated with a given maximum peak, can
be approximated as
\begin{equation}
\FAP(z) \lesssim M(z)\approx W e^{-z} \sqrt z, \quad W=\Delta f T_{\rm eff}
\label{prdgFAP}
\end{equation}
where $z$ is the periodogram maximum and $\Delta f$ is the adopted frequency range to scan.
Also, $T_{\rm eff}$ is an effective time-span of the data, which is proportional
to the weighted variance of $t_i$:
\begin{equation}
T_{\rm eff}^2 = 4\pi \left[ \frac{s_2}{s_0} - \left(\frac{s_1}{s_0}\right)^2 \right], \quad
s_k = \sum_{i=1}^N \frac{t_i^k}{\sigma_i^2}
\label{Teff}
\end{equation}
When dealing with periodograms that are tied to a particular data set in a heterogeneous
time series (as explained above), we should just restrict the summations in~(\ref{Teff}) to
the relevant subset.

The sign '$\lesssim$' in~(\ref{prdgFAP}) means that the function $M(z)$ in~(\ref{prdgFAP})
represents simultaneously an upper bound and an approximation for the $\FAP(z)$. The latter
approximation is asymptotic: for large $z$ (small $\FAP$) its error decreases. It is
important that the formula~(\ref{prdgFAP}) was originally constructed only for
strictly linear models with only the frequency of the probe sinusoidal signal allowed to be
non-linear. However, for non-linear but linearizable
models the approximation~(\ref{prdgFAP}) often should work in the asymptotic
sense, assuming $N\to\infty$ \citep{Baluev08b}. Therefore, possible deviations between the
simulated and analytic $\FAP$ curves, especially the ones that break the
inequality in~(\ref{prdgFAP}) may indicate the presence of a significant non-linearity. See
also additional discussion in \citep{Baluev14b}.

Likelihood-ratio periodograms can be easily defined for models involving red
noise. The significance levels for such periodograms can be constructed using the same
approach. Similarly to the white noise case, we may construct the quadratic Taylor
decomposition of the log-likelihood function~(\ref{loglik-corr}) (near its maximum). Such a
decomposition allows to transform all the formulae to a shape similar to the classic linear
regression task considered in \citep{Baluev08a}. In the end, it appears that the
formula~(\ref{prdgFAP}) is again asymptotically applicable even for models with correlated
(but still well linearizable) noise, with a slightly generalized value of $T_{\rm eff}$.
Namely, we should now substitute in~(\ref{Teff})
\begin{equation}
s_0 = \bmath u^{\rm T} \mathbfss V^{-1} \bmath u, \quad
s_1 = \bmath u^{\rm T} \mathbfss V^{-1} \bmath t, \quad
s_2 = \bmath t^{\rm T} \mathbfss V^{-1} \bmath t,
\label{s012}
\end{equation}
where $\mathbfss V$ is the compound noise covariance matrix of the RV data
(corresponding to the best fit), $\bmath u$ is a vector containing $N$
units, and $\bmath t$ is a vector containing all $t_i$. In the red-noise case we
should also satisfy the extra requirement $\Delta f \gg 1/\tau$ with $\tau$ being the noise
correlation timescale. Finally, when dealing with periodograms referring to individual
RV subsets of the heterogeneous time series (see above), we should
restrict the vectors $\bmath u$ and $\bmath t$ in~(\ref{s012}) to this
particular subset, but the matrix $\mathbfss V$ should remain intact.

The values of $T_{\rm eff}$ usually appear somewhat larger for the red noise, but it seems
that the difference in the predicted $\FAP$ levels often appears negligible (see
e.g. Fig.~\ref{fig_55Cnc_pow} discussed below).

\section{Testcase simulations}
\label{sec_simul}
We performed some Monte Carlo simulations revealing the effect of the jitter model
on various statistical distributions associated to the best fitting models. In this section
we consider relatively demonstrative examples involving RV data for several well-known
exoplanetary systems.

Originally we noticed the effects of the non-linearity in the additive jitter model when
considered the simulated distributions of periodograms. For large $N$, such periodograms
should asymptotically obey the Lomb-Scargle periodogram distribution derived
in \citep{Baluev08a}, provided that the RV curve model is well-linearizable. But
contrary to our expectations, the simulations show rather large deviations too often, even
for very large data sets and rather simple RV models. As we demonstrate below, in many
cases such deviations are owed to the non-linearity of the additive jitter model, rather
than to the `genuine' non-linearity of the RV model. We show that our regularized model of
the jitter can significantly improve the statistical behaviour of such periodograms.

Additionally to periodograms, we noticed that the additive noise model may significantly
exacerbate the non-linearity impact in the cases that involve a correlated Doppler
noise. We demonstrate below that our new regularized noise model may be very helpful
in such cases too.

\subsection{Doppler data for 51~Pegasi}
In the first testcase we consider the public radial velocity measurements of
51 Pegasi: $256$ ELODIE ones \citep{Naef04} and $153$ Lick ones \citep{Butler06}. We first
set a base model of these data. This includes a Keplerian almost
sinusoidal variation induced by the planet 51~Peg~\emph{b}
\citep{MayorQueloz95}, the long-term linear trend \citep{Butler06,Baluev08b}, and a
sinusoidal annual variation in the ELODIE data responsible for some instrumental drifts or
data reduction errors \citep{Baluev08b}. For the sake of symmetry between Lick and ELODIE,
we added a sinusoidal anuual variation to the model of the Lick data too, although contrary
to ELODIE its amplitude was not statistically significant. In these annual
variations the frequency was considered as a fixed
parameter, while the trigonometric coefficients were freely fittable. The compound
base model looks like
\begin{eqnarray}
\mu_{\mathcal H}(\btheta_{\mathcal H}, t) = c_{0i} + l t + a_i \cos(2\pi f_s t) + b_i \sin(2\pi f_s t) + \nonumber\\
+ K \{\cos[\omega+\upsilon(2\pi f t + \varphi,e) ]+e\cos\omega\}, \nonumber\\
\btheta_{\mathcal H} = \{c_{01},c_{02},l,a_1,b_1,a_2,b_2,K,f,\varphi,e,\omega\}
\label{51PegRV}
\end{eqnarray}
where $f_s=1$~year$^{-1}$ (fixed), and the last term of the model (on a separate line)
is the usual Keplerian signal due to 51~Peg~b. The index $i$ in~(\ref{51PegRV}) is equal to
either $1$ or $2$, depending on whether the model is computed for an ELODIE or Lick data
point. Basically this model contains only a couple of sinusoids (the Keplerian
signal has almost zero eccentricity and thus is almost sinusoidal) plus a linear
trend. Meanwhile, the number of the measurements in each of the datasets is very large, so
the RV curve model should be pretty linearizable. We would not expect that so simple model
can generate any non-linear overfit effects with so large time series. With this
RV curve model, we consider three different models of the Doppler noise: the classic
multiplicative one~(\ref{mult}), the legacy additive
ones~(\ref{add},\ref{addtrunc}), and the new regularized one~(\ref{wnreg}).

Apparently no pitfalls were encountered with the fitting of any of these noise models. The
best fitting values of the RV curve parameters are pretty expected and unremarkable, so we
do not detail them here. We only note that while the derived ELODIE best fitting
jitter value of $\sigma_\star= 6.5\pm 0.8$~m/s was well separated from zero, the Lick RV
jitter appeared consistent with zero within its uncertainty: $\sigma_\star^2=p=-0.5\pm
2.4$~(m/s)$^2$. The singularity of the additive noise model is located at the jitter value
of $-\min \sigma_{i,\rm meas}$. In the Lick data this minimum uncertainty is
$2.4$~m/s, corresponding to only a $2.2$-sigma separation between the derived jitter
estimation and the singularity. The average uncertainty value is
$\sigma_{\rm scale}=6.3$~m/s. Therefore, we may expect that the Lick data represent, due to
their negative jitter, a hidden source of the non-linearity and non-robustness of the fit,
even if the RV curve model itself looks well linearizable. On contrary, the ELODIE
data should not produce any such problems. For them we have $\sigma_{\rm scale}=7.3$~m/s
with almost all $\sigma_{i,\rm meas}$ being equal to $7$ or $8$~m/s (minimum is $7$~m/s).

Note that here we applied the modification to the likelihood function~(\ref{loglik})
that allows to perform a preventive bias reduction in the derived jitter
values \citep{Baluev08b}. Therefore, it is not the statistical bias that makes the
estimation of the Lick jitter close to zero and even negative. Actually, thanks to the
large number of the data, this bias correction appears rather small,
so the results would be qualitatively the same even if we used the classic
likelihood function~(\ref{loglik}).

\begin{figure*}
\includegraphics[width=0.45\textwidth]{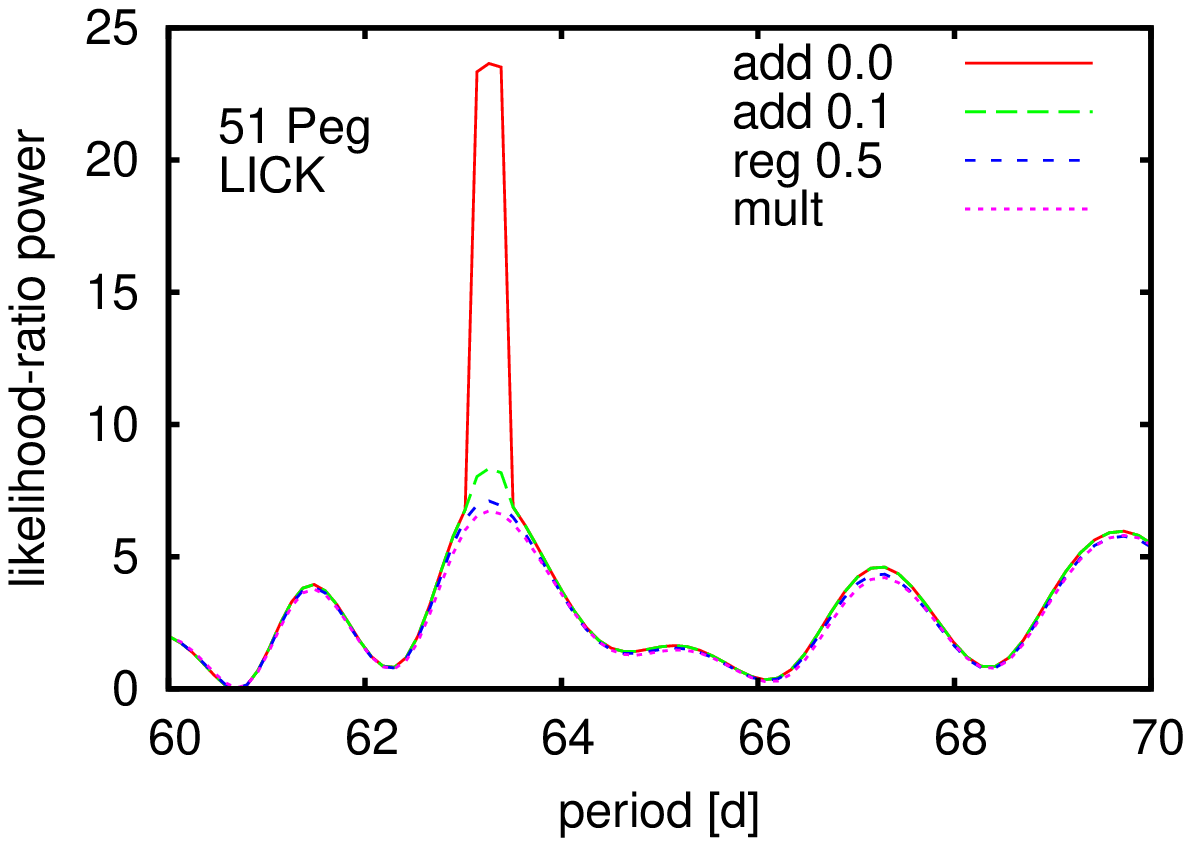}
\includegraphics[width=0.45\textwidth]{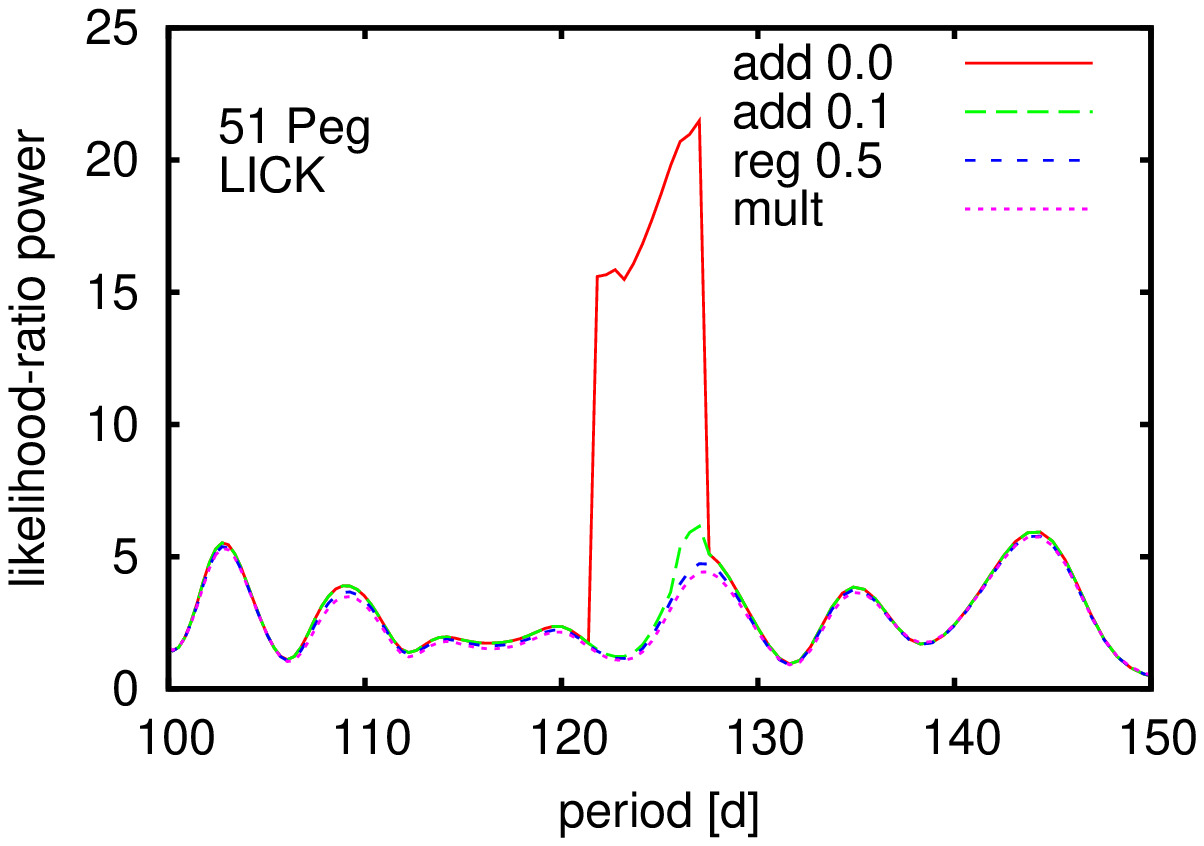}
\caption{Likelihood-ratio residual periodogram of the Lick
data for 51~Pegasi, remaining after removal of the contribution from the
planet 51~Peg~\emph{b}, linear RV trend, and systematic annual variation. Only two
narrow period ranges are shown to demonstrate bogus peaks generated by the
irregularities of the additive Doppler noise model. The curve labelled as `add
0.0' corresponds to the plain additive noise~(\ref{add}), `add 0.1' is for the truncated
additive noise~(\ref{addtrunc}) with the truncation parameter $a=0.1$, `reg 0.5' is for the
regularized noise model~(\ref{wnreg}) with the regularization parameter $a=0.5$, and `mult'
corresponds to the multiplicative noise model~(\ref{mult}).}
\label{fig_51Peg_pow}
\end{figure*}

Now we consider various likelihood-ratio residual periodograms of these RV data. First
of all, let us look at the Lick periodogram (Fig.~\ref{fig_51Peg_pow}). In these plots
we show two selected chunks in the period axis, containing the false peaks caused by the
irregularity of the additive model. We can see that the plain additive model~(\ref{add})
fails here, and the fitting algorithm is trapped in the singularity. The heights
of the bogus peaks should be actually infinite, but the fitting was stopped due to some of
the internal stopping criteria (likely, maximum number of iterations reached).
This periodogram contains many such bogus peaks, spanning a wide period range, and
basically rendering the periodogram itself useless. This behaviour is largerly corrected by
the truncated additive model~(\ref{addtrunc}) that effectively cuts the false peaks.

However, the periodogram for truncated additive model is still not always smooth
in the vicinities of the problematic periods. It is clear that
although the singularities themselves were cut off, the non-linearity that
always resides in the domain around a singularity is still not
eliminated and not suppressed. Although this non-linearity does not generate any obvious
undesired effects in this individual periodogram, we may suspect that it may introduce
certain biases when dealing with large randomized periodogram ensembles. This means that
various probabilistic estimations, like statistical uncertainties and significance
thresholds may get disturbed. For example, the average noise level
on the periodogram may be increased in comparison with what we would expect from a
well-linearizable case, resulting in a decreased detection efficiency.

Finally, our regularized noise model~(\ref{wnreg}) allows to obtain an entirely smooth
periodogram. Because we deal here with negative jitter (i.e. overestimated
instrumental errors), this periodogram stays close to the periodogram that was obtained for
the multiplicative model~(\ref{mult}). On contrary, if the best fitting jitter was
positive, the periodogram with regularized noise would be closer to the one with
additive noise. Therefore, our regularized noise model is able to automatically decide
whether the additive model is realistic or not for a particular data, and smoothly
select the suitable behaviour.

\begin{figure}
\includegraphics[width=84mm]{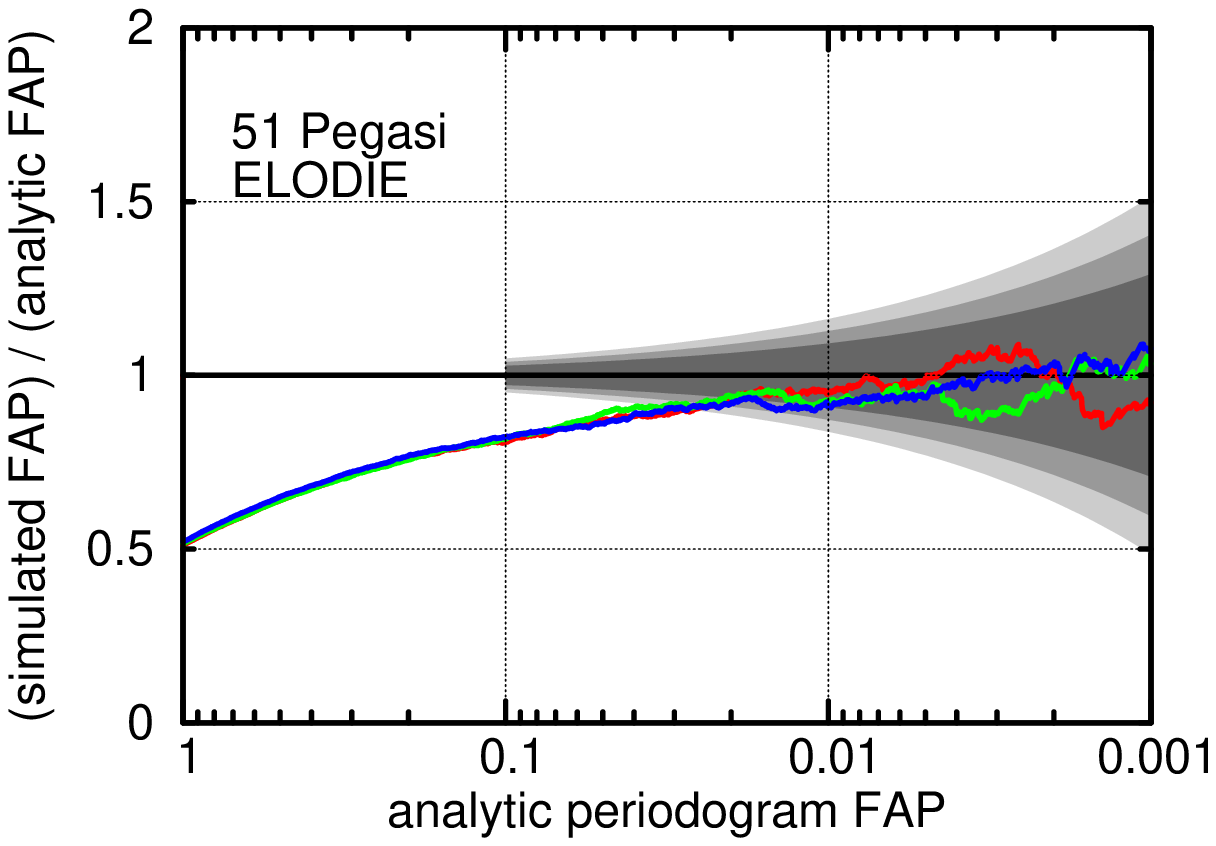}
\includegraphics[width=84mm]{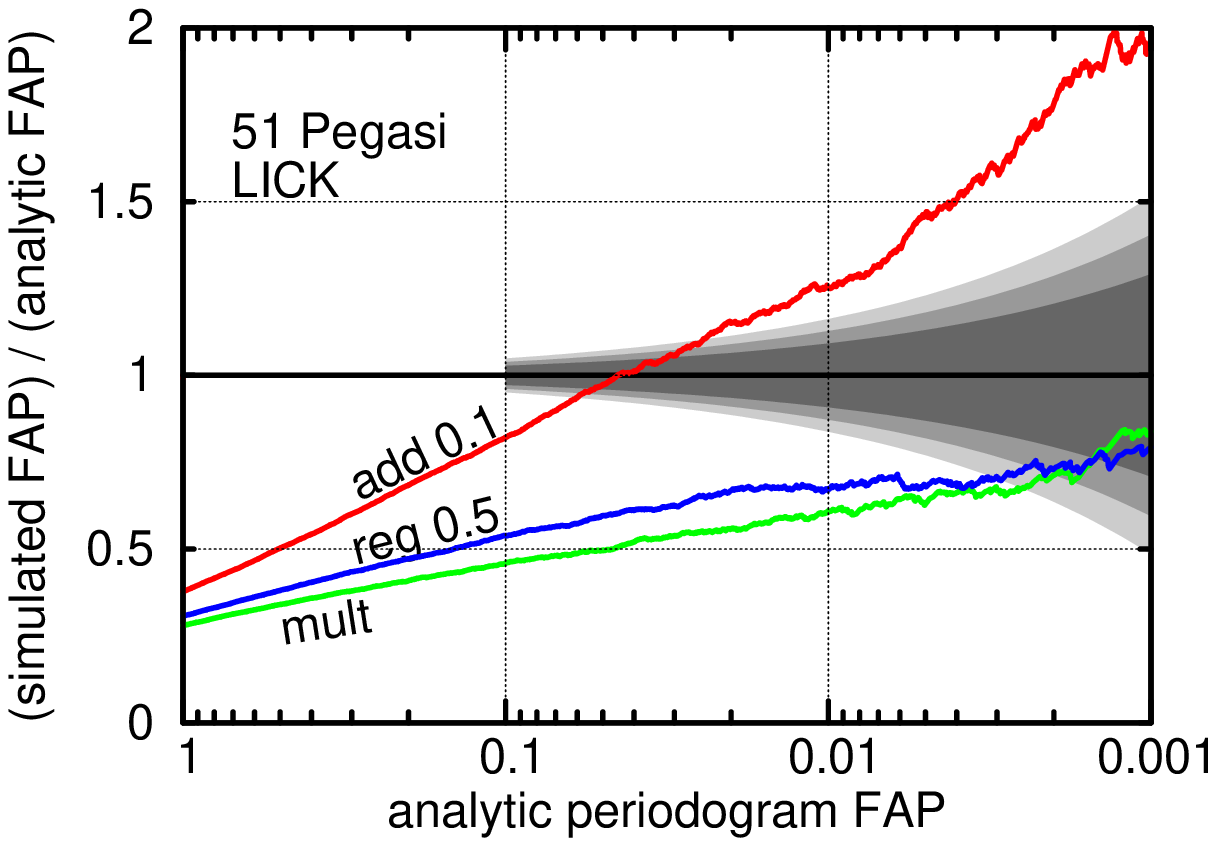}
\includegraphics[width=84mm]{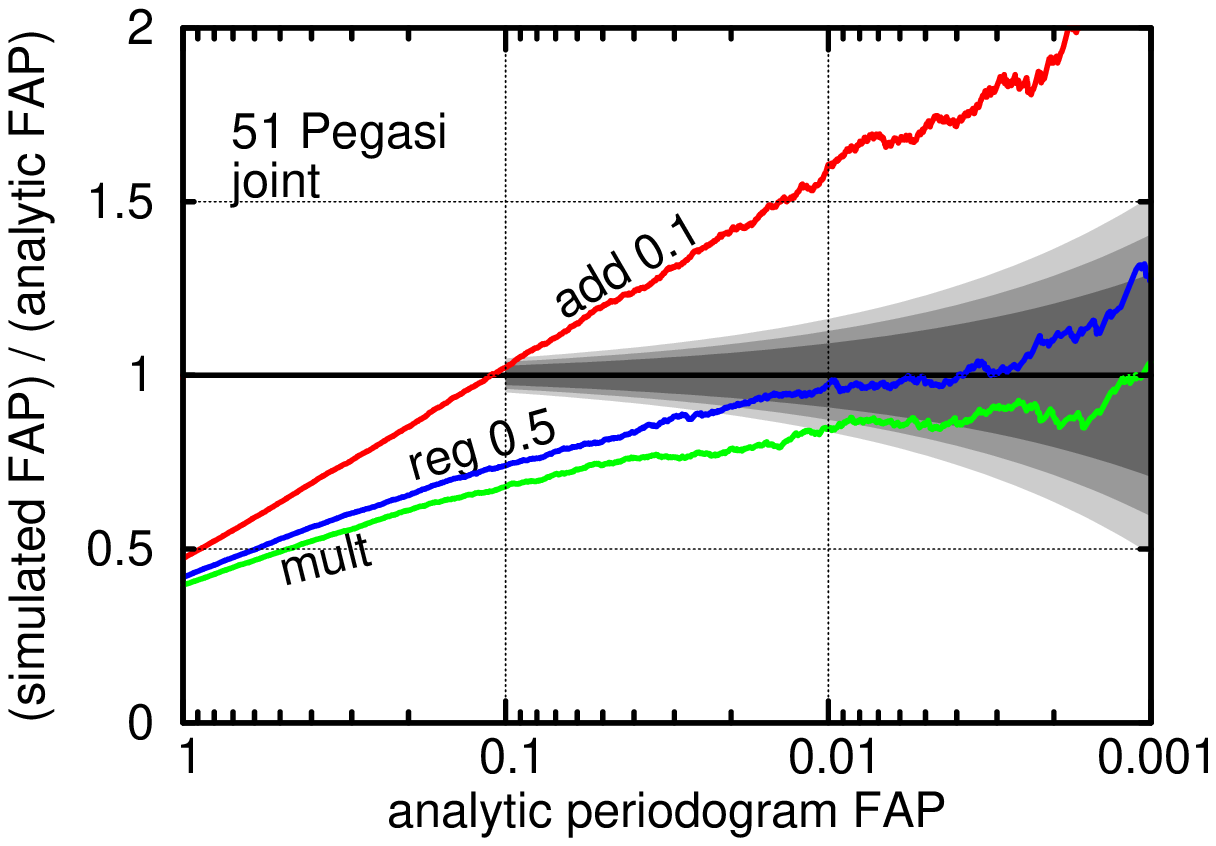}
\caption{Simulated periodogram false alarm probability in comparison with their analytic
approximation~(\ref{prdgFAP}). The three panels show the graphs of the
simulated/analytic $\FAP$ ratio for periodograms of the ELODIE data (top), Lick data
(middle), and joint data (bottom). The meaning of the curve labels is the same as in
Fig.~\ref{fig_51Peg_pow}. See text for a discussion.}
\label{fig_51Peg_fap}
\end{figure}

Now let us proceed to periodogram significance levels. For each of the three noise
models mentioned above we perform a Monte Carlo simulation
of the Lick, ELODIE, and joint periodograms in the frequency range from $0$ to
$1$~day$^{-1}$. Based on each of these $9$ simulations
we construct the empirical cumulative distribution function (CDF) of the periodogram
maximum, and compare this CDF with its analytic approximation~(\ref{prdgFAP}). The results
are shown in Fig.~\ref{fig_51Peg_fap}. In these graphs we plot the ratio $\FAP(z)/M(z)$ as
a function
of $M(z)$, for $9$ mentioned periodograms. Since Monte Carlo simulations infer some random
uncertainties that should be honoured, we also plot (in grayscale) the $1,2,3$-sigma
confidence ranges for this ratio, computed by means of the weighted Kolmogorov-Smirnov test
(see Appendix~\ref{sec_wKS}).

First of all, we can see that the ELODIE periodograms (first panel) do not
show any signs of non-linearity with any of the noise models. In fact, these plots
only reveal that the formula~(\ref{prdgFAP}) behaves exactly as it was supposed to
behave in the linear case: it represents an asymptotic approximation to the $\FAP$
when $\FAP\to 0$, simultaneously serving as an upper limit for larger $\FAP$s. Also, we may
note that all three $\FAP$ curves generated by different noise models almost coincide with
each other within the Monte Carlo uncertainties. This is because of the small spread
of $\sigma_{i,\rm meas}$ for the ELODIE RV data: in fact, almost all
of these $\sigma_{i,\rm meas}$ are equal to either $7$~m/s or $8$~m/s. In this case we have
all $\tau_i$ in~(\ref{wnreg},\ref{wnmultadd}) close to unit, implying that all three noise
models are practically equivalent to each other. In this case the weights of the
observations become almost equal for any of these models, and we just need to assess the
unknown common noise variance.

In the second panel of Fig.~\ref{fig_51Peg_fap} the results for the Lick
periodograms are shown. In this case the $\FAP$ curve for the truncated additive model does
not obey the analytic model~(\ref{prdgFAP}). It is remarkable and rather disappointing that
even the inequality in~(\ref{prdgFAP}) got broken: the simulated $\FAP(z)$ may be even
twice as large as $M(z)$. This indicates that Lick periodogram with additive noise
possesses an increased noise level, in comparison with what we might expect from a
linearizable fit. As we suspected, the truncated modification~(\ref{addtrunc}) is unable to
suppress this non-linearity. However, our regularized model is clearly able to suppress it,
making the analytic formula~(\ref{prdgFAP}) useable again. The results for the periodogram
of the joint ELODIE+Lick data (third frame in Fig.~\ref{fig_51Peg_fap}) look similar to the
Lick case.

The summary is that the additive RV noise model, including its truncated
version, often generates an excessive noise on the periodograms in the cases
when the estimated jitter value is negative or consistent with zero (taking into account it
uncertainty). Obviously, this noise excess may only decrease the detection efficiency, and
increase the statistical uncertainty of the usual RV curve parameters. This noise excess is
generated by the non-linearity of the RV noise model only, and thus it can appear even when
the RV curve model is very simple and does not show any significant non-linearity in
itself.

Fortunately, this non-linearity can be efficiently suppressed by replacing the
additive noise model with the regularized one. Moreover,
with the regularized noise model we noticed up to $\sim 30$ per cent speed-up of the
computations. The likely explanation is that
to fit a well-linearizable model a smaller number of Levenberg-Marquardt iterations is
needed than to fit a highly non-linear model (in the ultimate limit of
strict linearity even a single iteration would be enough).

\subsection{Doppler data for GJ~581}
\label{sec_GJ581}
The RV data for this star contain a significant correlated noise component
with a correlation timescale of $\sim 10$~d \citep{Baluev13a,Tuomi12}. Here we
revist the results of \citep{Baluev13a} in view of the improved noise models
introduced above.

In \citep{Baluev13a} we used the approach described in Sect.~\ref{sec_rednoise} above, but
it was unintentionally allowed to have $p_{\rm red}$ negative, unless it
breaks the positive definiteness of the noise covariance matrix $\mathbfss V$. Also,
we used the plain (not even truncated) additive model of the white noise in this case.
As we now demonstrate, these issues caused significant corrupting effect on some
of the Monte Carlo simulations from \citep{Baluev13a}. Although the Monte Carlo trials
generating $p_{\rm red}<0$ were seldom, they triggered drastically large values of the
maximized likelihood function, introducing significant distortions
in various simulated distributions.

\begin{figure*}
 \includegraphics[width=0.45\textwidth]{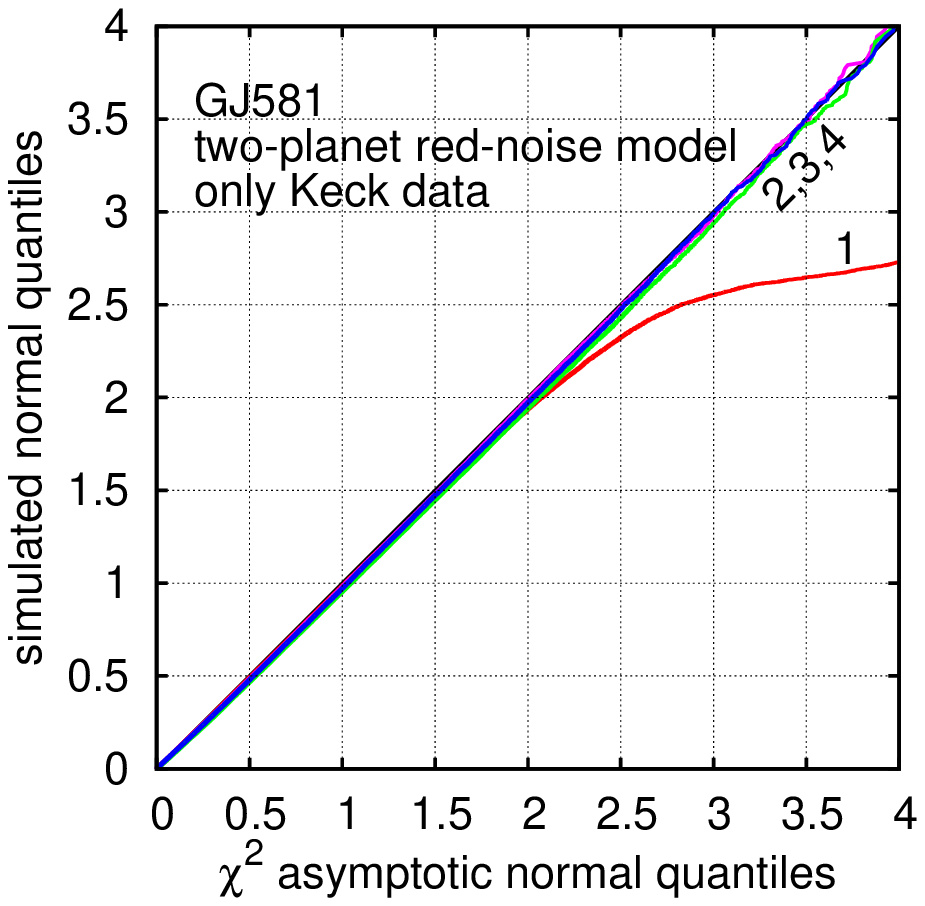}
 \includegraphics[width=0.45\textwidth]{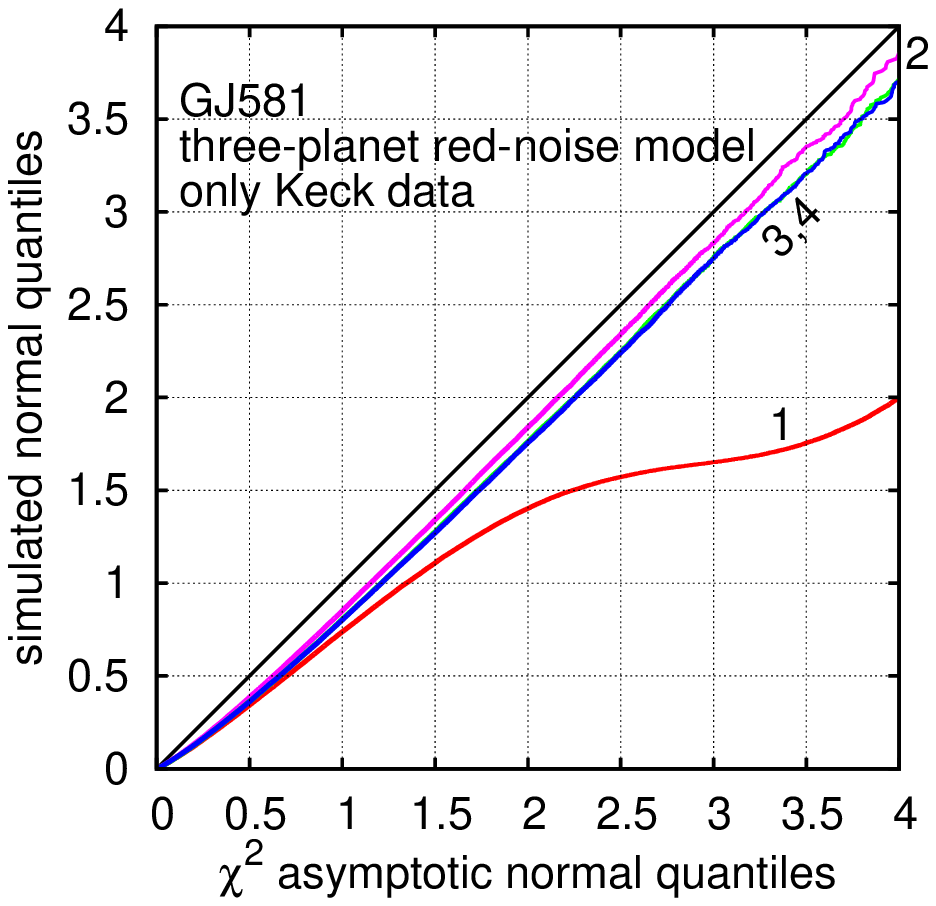}
\caption{Simulated distribution of the modified likelihood-ratio
statistic $\tilde Z$ of the GJ~581 fits, compared with its asymptotic $\chi^2$ distribution
in terms of their normal quantiles (the "$n$-sigma" significance
levels). The statistic $\tilde Z$ is defined in \citep{Baluev08b}. The null hypothesis
$\mathcal H$ constrains the set of the fit parameters to a single fixed point, while the
alternative one $\mathcal K$ treats all the parameters free. The curves labelled with an
index $1$--$4$ correspond to the following noise models: ($1$) the plain
additive model applied in \citep{Baluev13a}, ($2$) the truncated
additive model described in Sect.~\ref{sec_GJ581}, ($3$) the model
with truncated red and regularized white noise, and ($4$) the model with truncated red and
multiplicative white noise. Number of Monte Carlo simulations is $10^5$ for all the cases.}
\label{fig_GJ581_mc}
\end{figure*}

This is shown in Fig.~\ref{fig_GJ581_mc}, where we compare simulated distributions
of the likelihood-ratio statistic for different noise models. The curves labelled with
``1'' are for the plain model~(\ref{rednoise}) and were taken from
\citep{Baluev13a}. The curves labelled with ``2'' correspond
to the model where the white noise component is expressed by the truncated
model~(\ref{addtrunc}) with $a=0.1$, and the red noise is also expressed by a ``truncated''
model, in which negative $p_{\rm red}$ are disallowed. The curves labelled
with ``3'' and ``4'' differ from the second case only in the model
of the white noise component, which is set to the regularized with $a=0.5$ or to the
multiplicative one, respectively. In the left panel of Fig.~\ref{fig_GJ581_mc} we show the
simulations done for the two-planet RV curve model, and in the right panel we include three
planets (this is the maximum number of planets that can be detected in this system
with only the Keck RV data). We can see that in both the cases, the curve ``1'' is
very different from the others, indicating that this simulation was significantly affected
by the non-linearity that emerges for negative $p_{\rm wht}$ or $p_{\rm red}$.

Similarly to the 51~Pegasi testcase, this non-linearity results in reduced
significance estimations and increased statistical uncertainties. Fortunately, it can be
suppressed by choosing another noise model. Therefore, this is in fact a
``fake'' non-linearity caused by the bad mathematical behaviour of an incarefully selected
noise model. Here we do not spot any remarkable difference between the truncated additive,
regularized, and multiplicative models of the white noise component,
although the difference with the plain model~(\ref{rednoise}) is obvious. A detailed
investigation revealed that in the two-planet case both types of the singularity
with $p_{\rm red}<0$ and with $p_{\rm wht}<0$ may appear
in the simulations (although not simultaneously of course), while for the three-planet
model the Monte Carlo trials with $p_{\rm red}<0$ do not occur. We may conclude that
in practice any of the two noise terms in~(\ref{rednoise}) can raise irregularity issues.

Since we now have an efficient method to suppress the non-linearity of these red-noise
fits, we used it to recalculate the significance of the fourth planet in the system,
GJ~581~\emph{d}. In \citep{Baluev13a} this significance was estimated by a rather marginal
value of $2.1$-sigma at most. Given the simulation results discussed above, we may
expect that this significance may be underestimated due to the poor quality of the RV noise
model, and in true this planet can be confirmed with a greater confidence. We re-run the
periodogram simulations from \citep{Baluev13a} assuming the regularized model~(\ref{wnreg})
for the white component and constraining the red jitter parameter $p_{\rm red}$ to
only positive values. The correlation shape function $\rho(x)$ was adopted exponential
again.

\begin{table*}
\caption{Detection significance levels for the
planet GJ~581~\emph{d}}
\label{tab_GJ581}
\begin{tabular}{lccc}
\hline
origin of the estimation & only HARPS data & joint HARPS+Keck data, & joint HARPS+Keck data,  \\
                         &                 & ``shared'' red noise   & ``separated'' red noise \\
\hline
Monte Carlo estimations from \citep{Baluev13a} & $1.8$-sigma & $2.1$-sigma & $1.3$-sigma \\
Monte Carlo estimations from the present work  & $2.1$-sigma & $2.6$-sigma & roughly $3.6$-sigma \\
Analytic formulae~(\ref{prdgFAP},\ref{Teff},\ref{s012}) & $2.2$-sigma & $2.7$-sigma & $3.6$-sigma \\
\hline
\end{tabular}\\
\small Normal quantiles corresponding to the relevant periodogram $\FAP$s are given. The
``shared'' and ``separated'' red noise models correspond to the cases when the
parameters for the Keck and HARPS red noise are bound with each other or are
fitted independently, see \citep{Baluev13a} for details. These results do not take
into account the recent \citet{Robertson14} work.
\end{table*}

We summarize our results in Table~\ref{tab_GJ581}. We can see that now the detection
significance levels of the planet~\emph{d} are remarkably increased. Moreover, the new
Monte Carlo estimations are now in a much better agreement with the analytic approximations
that assume a linearizable task. The simulated significances are slightly larger
than the analytic ones, formally breaking the $\FAP$ inequality~(\ref{prdgFAP}), still
indicating the presence of some residual non-linearity in the models. However, the
deviation is now modest and pretty negligible. This indicates that the detection
of the GJ~581~d signal is now statistically robust.

Also, our updated results do not call back the conclusion of \citet{Baluev13a} that adding
the correlated noise component drastically reduces the periodogram peak corresponding to
the planet~\emph{d}. In fact, after applying our regularized model the height of this peak
in the periodograms \emph{of the real data} changes negligibly in comparison with what
we had in \citep{Baluev13a}. All the distortions are living in some fraction
of the simulated periodograms, increasing the periodogram noise. Also, the conclusions
of \citet{Baluev13a} concerning the more speculative planet candidates
\emph{f} and~\emph{g} remain intact: the relevant periodogram peaks just disappear after
turning on the RV noise correlations in the model. Here we further confirm
that these planets likely do not exist or at least not supported by the current RV data.

Although our new reanalysis of the GJ~581~d signal suggests that it appears formally
more significant than in \citep{Baluev13a}, we still do not insist on its existence. During
the refereeing of the present paper, a new work by \citet{Robertson14} appeared, where the
authors claimed that this signal owes to the stellar activity manifesting itself
in the Doppler data for GJ~581. However, they demonstrated that the actual correlated noise
in the GJ~581 HARPS data is not stationary, likely showing some cyclic variability in time.
Contrary to this, our analysis, both in \citep{Baluev13a} and here, assumes a constant
activity level and constant red noise characteristics. If the actual red noise is varying,
its reduction in the constant-level approximation might appear incomplete. This means that
the significance estimates in Table~\ref{tab_GJ581} should decrease back
when the variations of the red noise characteristics are taken into account. It seems that
GJ~581 is in this regard reminiscent of the 55~Cnc case considered below. However, in this
paper we had no aim to consider the GJ~581 data in so deep details. Here we only
used its data as a testcase to demonstrate the non-linearity
of the white and red noise models and verify that our regularization scheme indeed
allows to suppress most of this non-linearity.

\subsection{Doppler data for HD~82943}
Of course, our regularized noise model is not a magic stick that would
suppress the non-linearity everywhere it is applied. We must remember that in some cases
the non-linearity of the fit genuinely belongs to the model of the RV curve, rather than of
the RV noise. Such an example is offered by the HD~82943
system \citep{Tan13,BaluevBeauge14}. The regularized noise model was designed to
handle more gracefully the cases in which the estimated jitter is consistent with zero
or even negative. However, in the HD~82943 all
jitter estimations \citep[see][]{BaluevBeauge14} are positive and well separated from zero.
For example, \citet{BaluevBeauge14} present the simulation of the periodogram of the
Keck data for this star (see their Fig.~12). This simulation shows significant deviation
from the analytic prediction~(\ref{prdgFAP}), owing to the fit non-linearity. Now we
recomputed this simulation based on the regularized noise model~(\ref{wnreg}), but
no visible change was revealed. In this case, the non-linearity is caused by the
traditional sources: the complexity of the RV curve model with a relatively small number of
the Keck RV observations.

\section{Correlated Doppler noise in the 55~Cancri RV data}
\label{sec_55Cnc}
\citet{Nelson14} presented a thorough investigation of the updated RV data for
this star. This investigation included the latest Keck \citep{Nelson14},
Lick \citep{Fischer14}, HET (Hobby-Eberly Telescope, \citealt{Endl12}), and HJST (Harlan J.
Smith Telescope, \citet{Endl12}) measurements. Although many subtle effects were considered
in the analysis by \citep{Nelson14}, e.g. a correlation of Doppler noise over a single
night due to stellar oscillations, they did not address the possibility of
RV noise correlations at the timescale of a few days or more. They only stated (without an
explanation) that the assumption of uncorrelated RV errors ``is usually an excellent
approximation when observations are separated by one or more days''. But as we discussed in
Sect.~\ref{sec_rednoise}, this assumption may appear wrong, because at the precision level
of $\sim 1$~m/s we frequently deal with RV correlations on the average time scale of $\sim
10$~d.

Here we undertake an attempt to address this issue of Doppler noise correlations
over the timescales greater than $1$~d, based on our new
approach involving regularized noise models. We perform a reanalysis of the above-mentioned
public RV measurements from Keck, Lick, HET, and HJST, also adding to them the
ELODIE dataset from \citep{Naef04}, which \citet{Nelson14} did not include. Following to
\citet{Nelson14,Fischer14}, we separate the Lick dataset into several subsets,
acquired using different Dewar, according to the information given in the
\citet{Fischer14} Lick RV database. We need to note that the Lick dataset used
by \citet{Nelson14} is a bit larger than the one published \citet{Fischer14}. It seems that
the difference is due to the treatment of the RV measurement series that were obtained in a
single night. Apparently, \citet{Fischer14} replace such series by some average value,
because these data do not contain so close $t_i$, while \citet{Nelson14} deal the full
(unpublished) data, treating the correlations inside the series close data points in two
distinct ways (their ``Case 1'' and ``Case 2''). Here we only have access to the
public \citet{Fischer14} data. From the discussion by \citet{Nelson14} we learned some
additional information about the \citet{Fischer14} RV data. The starting group of $14$
RV measurements that are labelled with Dewar code 8 is actually related to
an ``early Doppler era'' and should be treated separately from the ending group with
formally the same Dewar code. Also, it appears that these early-era data should be
separated in two even smaller subsets ($4+10$ points) that refer to different Dewar and may
have different statistical characteristics. However, due to the small number
of these observations we cannot model their jitter separately: such an attempt
frequently leads to parametric degeneracies. We therefore decided to share a single jitter
model between these two RV subsets, still allowing them to have different RV offsets. We
must note that these $14$ RV measurements anyway have a small effect of the entire
RV fit due to their small number and relatively large uncertainties.

\begin{figure}
 \includegraphics[width=84mm]{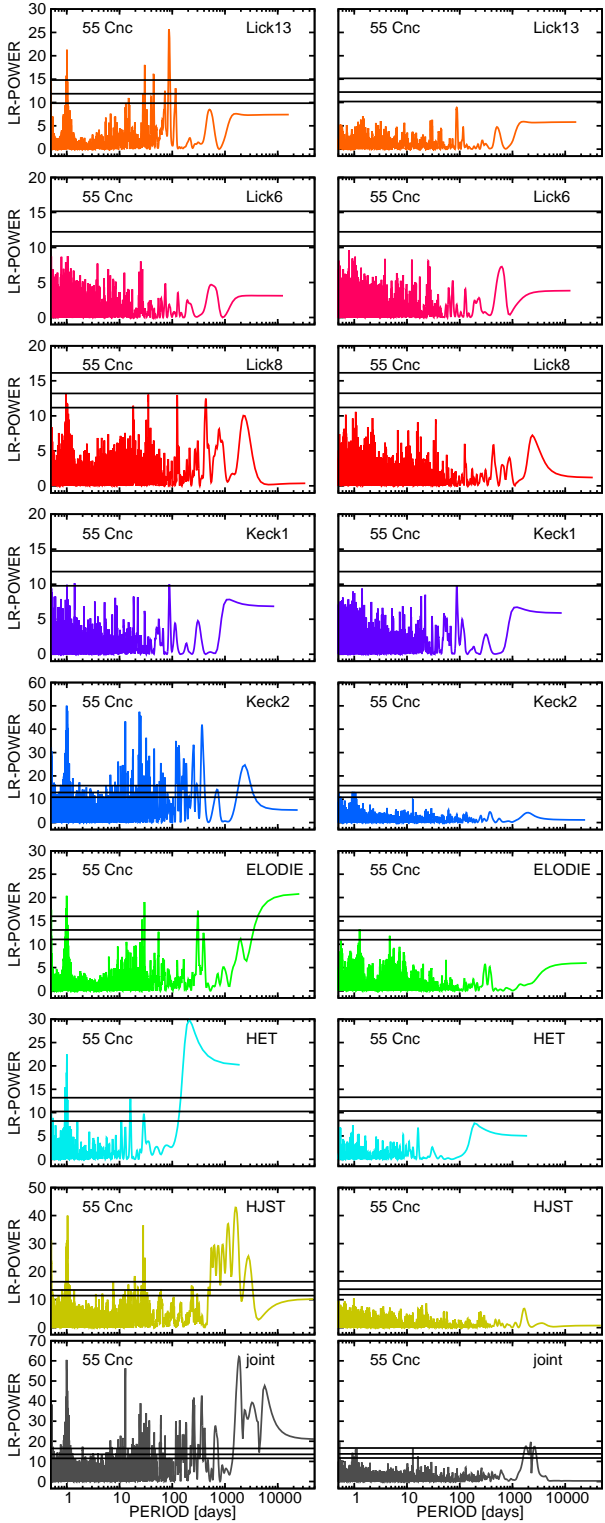}
\caption{Residual likelihood-ratio periodograms for $8$ individual
RV datasets and for the joint RV dataset of 55~Cnc. Left column is for the
purely white model of the RV noise with independent jitters, right column corresponds to
the correlated (white+red) Doppler noise model, assigned separately for all individual
datasets except for the `Lick6' and `Keck1' ones, which still preserve
only the white component. The white noise is approximated by the regularized noise
model. The RV curve model includes contributions from all $5$ known planets treated in the
unperturbed (Keplerian) framework.}
\label{fig_55Cnc_pow}
\end{figure}

At first, we considered the unperturbed (multi-Keplerian) model of the RV curve, involving
all $5$ known planets orbiting 55~Cnc, and approximating the RV noise in the
above-discussed datasets by the regularized model~(\ref{wnreg}). These
periodograms based on the regularized white noise model are shown in the left column
of Fig.~\ref{fig_55Cnc_pow}. For each of these periodogram we also plot a triplet of
$1$, $2$, and $3$-sigma noise levels computed using the analytic formula~(\ref{prdgFAP}).

We can see that they look very differently for different datasets. In the most cases we can
see a clearly non-uniform frequency distribution, indicating that the Doppler noise in not
white.\footnote{With the logarithmic scale of the period axis used
in Fig.~\ref{fig_55Cnc_pow} the white noise should appear monotonically growing to the
short-period end of the axis. If such a pattern is not fulfilled, the noise is not white.}
Only the Lick Dewar~6 subset and the Keck-1 subset are consistent with the white
noise. Simultaneously they agree well with the predicted noise levels. The other RV subsets
show remarkable excess power at the periods $>10$~d and in a narrow period range
around $1$~d. This is a typical picture that the so-called red noise
generates: the power excess at long periods is aliased to the vicinity of the $1$~d period
due to diurnal gaps existing in the RV data. Simultaneously, these periodograms do not
comply with the predicted noise levels expected for the white noise. We can see many peaks
that rise well above even the three-sigma level, but these peaks are different for
different datasets, indicating some noise effects rather than an actual RV periodicity.

We modelled this red noise separately for each RV dataset, excluding the Lick-6 and Keck-1
ones (they preserved only white components). The model of the red part
of the noise was analogous to the one used for the GJ~581 case (see above), constraining
$p_{\rm red}>0$. The white part noise was modelled by the regularized model again. The
cross-correlations between different RV subsets were set to zero. The periodograms based on
such compound noise model are plotted in the right column of Fig.~\ref{fig_55Cnc_pow}.
We can see that the red noise is largerly suppressed in these periodograms. In the
most cases the periodograms become consistent with the predicted noise levels,
indicating an almost complete reduction of the red noise. Only the Keck-2 and ELODIE
data still contain some remaining non-whiteness and marginally significant peaks.

\begin{figure*}
 \includegraphics[width=0.75\textwidth]{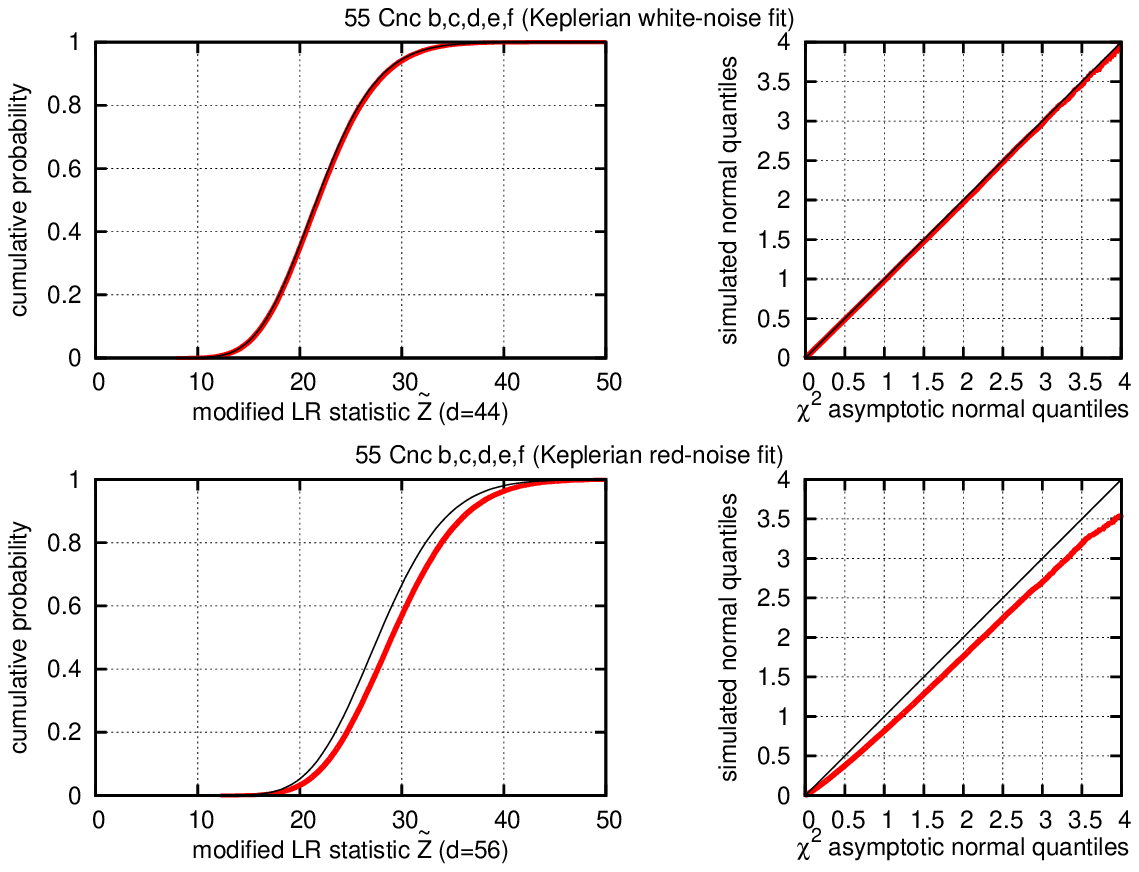}
\caption{Simulated distribution of the modified likelihood-ratio
statistic $\tilde Z$ of the 55~Cnc fit, compared with its asymptotic $\chi^2$ distribution.
The cumulative distributions of $\tilde Z$ is shown in the
left panels, while the right panels compare their normal quantiles. See also the notes
of Fig.~\ref{fig_GJ581_mc}. Here we only use the model
with the truncated red and regularized white noise.}
\label{fig_55Cnc_mc}
\end{figure*}

To verify that the non-linearity of the red-noise model is small enough and our statistical
methods a legal here, we perform Monte Carlo simulations of the likelihood-ratio statistic.
These are done similarly to the GJ~581 case above, and the results are shown
in Fig.~\ref{fig_55Cnc_mc}. We can see that in the white-noise
case the agreement between the simulated and asymptotic $\chi^2$ distribution is very good.
For the red-noise model only some minor deviation appears. Therefore, we believe that both
the RV curve and RV noise models do not generate significant non-linearity in the fit, and
our results should be pretty reliable in terms of their statistical robustness.

\begin{figure}
 \includegraphics[width=84mm]{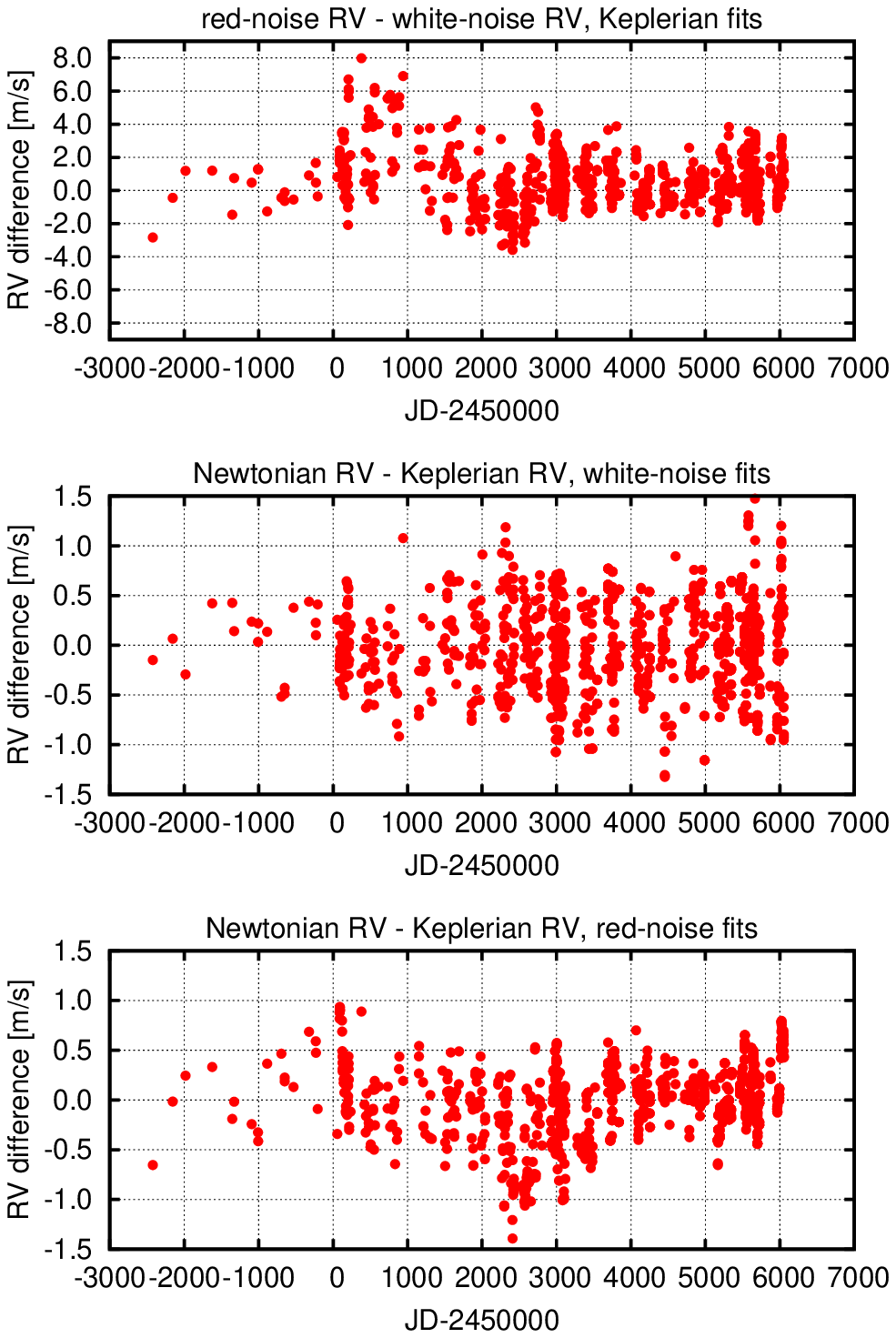}
\caption{Differences between best fitting RV curve models of 55~Cnc, plotted
at the discrete times of the available RV observations as points. The models used to obtain
the fits under comparison are labelled on top of each panel. In each of the cases all five
known planets were included.}
\label{fig_55Cnc_rsd}
\end{figure}

We also consider the impact of planet-planet Newtonian perturbations on the fit.
We consider best fits based on either Keplerian or Newtonian (coplanar edge-on) model of
the RV curve, and calculate the difference between the inferred RV curve models.
Additionally, we compute the difference between the best fitting RV curves for the white
and red noise models (neglecting Newtonian perturbations). These differences are plotted in
Fig.~\ref{fig_55Cnc_rsd}. We can see that Newtonian perturbations have typical value
of only $0.5$~m/s over the observation time span, significantly smaller than the
RV uncertainties. Moreover, we can see that the impact
of the red noise exceeds the Newtonian perturbations roughly by the factor
of $\sim 4$. This means that the RV data available for 55~Cnc are still unable to reveal
any signs of Newtonian perturbations in this planetary system.
Therefore, these Newtonian perturbations have only a negligible effect on the periodograms
and simulations discussed above. Moreover, it may even appear that we wrongly interprent
some phantom red-noise variation as a hint of Newtonian perturbations. From the other side,
\citep{Nelson14} show that the fit accuracy is enough to have a significant difference
between the osculating and averaged orbital elements, exceeding the estimation uncertainty
by the factor of a few. These conclusions do not contradict to each other: the presence of
detectable offsets between the osculating and averaged orbital elements does not
necessarily imply the presence of detectable perturbations in the RV curve (except for a
secular linear drift in the planetary mean longitudes that act like a period biasing).

\begin{table*}
\caption{Best fitting parameters of the 55~Cnc: five-planet Newtonian edge-on model with
red noise at JD2454000.}
\label{tab_55Cnc}
\begin{tabular}{llllll}
\hline
\multicolumn{6}{c}{planetary orbital parameters and masses}  \\
                       & planet b         & planet c       & planet d       & planet e       & planet f       \\
$P$~[d]                & $14.65152(15)$   & $44.4175(73)$  & $4825(39)$     & $0.7365515(15)$& $262.00(51)$   \\
$K$~[m/s]              & $71.40(29)$      & $10.18(33)$    & $48.29(80)$    & $5.97(17)$     & $4.87(43)$     \\
$e$                    & $0.0034(32)$     & $0.020(31)$    & $0.019(13)$    & $0.040(27)$    & $0.305(75)$    \\
$\omega$~[$^\circ$]    & $98(53)$         & $51(84)$       & $44(31)$       & $72(42)$       & $166(15)$      \\
$\lambda$~[$^\circ$]   & $61.44(22)$      & $228.4(2.0)$   & $291.78(81)$   & $134.3(1.6)$   & $185.6(5.1)$   \\
$i$~[$^\circ$]         & \multicolumn{5}{c}{$90$ (fixed)} \\
$M$~[$M_{\rm Jup}$]    & $0.8306(33)$     & $0.1714(55)$   & $3.878(68)$    & $0.02561(73)$  & $0.141(12)$    \\
$a$~[AU]               & $0.11522725(79)$ & $0.241376(26)$ & $5.503(30)$    &$0.015690907(21)$&$0.7880(10)$   \\
\hline
\multicolumn{6}{c}{parameters of the data sets} \\
                           & Lick early 1   & Lick early 2   & Lick 13        & Lick 6         & Lick 8         \\
$N$                        & $4$            & $10$           & $60$           & $54$           & $162$          \\
$T$~[d]                    & $797$          & $824$          & $1623$         & $1257$         & $3336$         \\
$T_{\rm eff}$~[d]          & $951$          & $912$          & $1323$         & $1343$         & $3466$         \\
$\sigma_{\rm scale}$~[m/s] & $8.559$        & $8.188$        & $1.941$        & $2.207$        & $1.493$        \\
white $\sigma_\star$~[m/s] & $-1.5(8.3)$    & $-1.5(8.3)$    & $0.6(1.6)$     & $4.12(51)$     & $3.69(47)$     \\
red   $\sigma_\star$~[m/s] & -              & -              & $7.6(1.2)$     & -              & $6.30(74)$     \\
$\tau$~[d]                 & -              & -              & $10.5(5.3)$    & -              & $15.6(6.3)$    \\
$c$~[m/s]                  & $9.5(4.3)$     & $29.9(3.6)$    & $-0.3(1.9)$    & $3.9(1.3)$     & $-1.2(1.0)$    \\
r.m.s.~[m/s]               & $9.047$        & $7.781$        & $8.077$        & $5.003$        & $7.016$        \\
\\
                           & Keck 1         & Keck 2         & ELODIE         & HET            & HJST           \\
$N$                        & $24$           & $469$          & $48$           & $131$          & $212$          \\
$T$~[d]                    & $858$          & $2390$         & $2571$         & $190$          & $4728$         \\
$T_{\rm eff}$~[d]          & $886$          & $2553$         & $2805$         & $219$          & $5644$         \\
$\sigma_{\rm scale}$~[m/s] & $1.294$        & $0.952$        & $7.231$        & $3.092$        & $2.735$        \\
white $\sigma_\star$~[m/s] & $3.31(55)$     & $1.310(79)$    & $-5.85(62)$    & $1.92(50)$     & $2.86(30)$     \\
red $\sigma_\star$~[m/s]   & -              & $3.27(26)$     & $14.3(2.7)$    & $5.4(1.0)$     & $5.7(1.0)$     \\
$\tau$~[d]                 & -              & $3.52(89)$     & $118(64)$      & $13.3(6.1)$    & $94(52)$       \\
$c$~[m/s]                  & $-30.9(1.1)$   & $-33.33(72)$   & $27267.5(4.6)$ & $28396.4(2.1)$ & $-22573.0(1.5)$\\
r.m.s.~[m/s]               & $3.497$        & $3.530$        & $16.278$       & $5.951$        & $6.342$        \\
\hline
\multicolumn{6}{c}{general characteristics of the data and fit}   \\
$N$                        & \multicolumn{5}{c}{$1174$}  \\
$T$~[d]                    & \multicolumn{5}{c}{$8476$}  \\
$T_{\rm eff}$~[d]          & \multicolumn{5}{c}{$5593$}  \\
$\tilde l$~[m/s]           & \multicolumn{5}{c}{$3.796$} \\
$d$                        & \multicolumn{5}{c}{$56$}    \\
\hline
\end{tabular}\\
The parameters have the following meaning: orbital period $P$, RV semiamplitude $K$,
eccentricity $e$, pericenter argument $\omega$, mean longitude $\lambda$, derived planet
mass $M$ and semimajor axis $a$. The osculating orbital elements are in the Jacobi
reference frame described in \citep{Baluev11}, but the innermost planet was involved
neither in the N-body integration, nor in the construction of the Jacobi reference
frame. The values of $M$ and $a$ were derived assuming the common orbital inclination of
$i=90^\circ$ and the mass of the star $M_\star = 0.95 M_\odot$. The uncertainty of
$M_\star$ was not included in the uncertainties of the derived values. The parameter $c$ is
the constant RV offset, and $\sigma_\star$ is the estimated magnitude of the white
or red RV jitter (the ``Lick pre1'' and ``Lick pre2'' datasets share the same jitter). The
values $\tau$ represent the correlation timescale of the red noise. The white jitter
is approximated by the regularized model with the specified jitter scale
factor $\sigma_{\rm scale}$ (see text). The quantities $T$ and $T_{\rm eff}$ represent a
dataset usual time-span and the effective time-span needed
to compute periodogram noise levels (see text). The goodness of the fit $\tilde l$ is tied
to the modified likelihood function as explained in \citep{Baluev08b}. The integers $N$ and
$d$ are the number of observations and of the free parameters, respectively.
\end{table*}

However, considering all pro and contras, we prefer to deal with the osculating elements
and Newtonian model, and we give this final 55~Cnc fit in Table~\ref{tab_55Cnc}.
This fit involves the correlated RV noise model and Newtonian coplanar edge-on model of the
RV curve. Comparing this fit with the fit by \citet{Nelson14}, we do not
spot any significant difference, except for the planet~\emph{f} eccentricity, $e_f$.
We obtain $e_f\approx 0.3$, while \citet{Nelson14} give an almost zero value. This
change appears due to the red-noise model. More detailed investigation shows that this
$e_f$ is still consistent with zero: we obtain only $\sim 0.5$-sigma deviation according to
the relevant likelihood-ratio statistic. In fact, this eccentricity appears very
poorly determined, offering only an upper limit of
$e_f\lesssim 0.5$.

\begin{figure}
 \includegraphics[width=84mm]{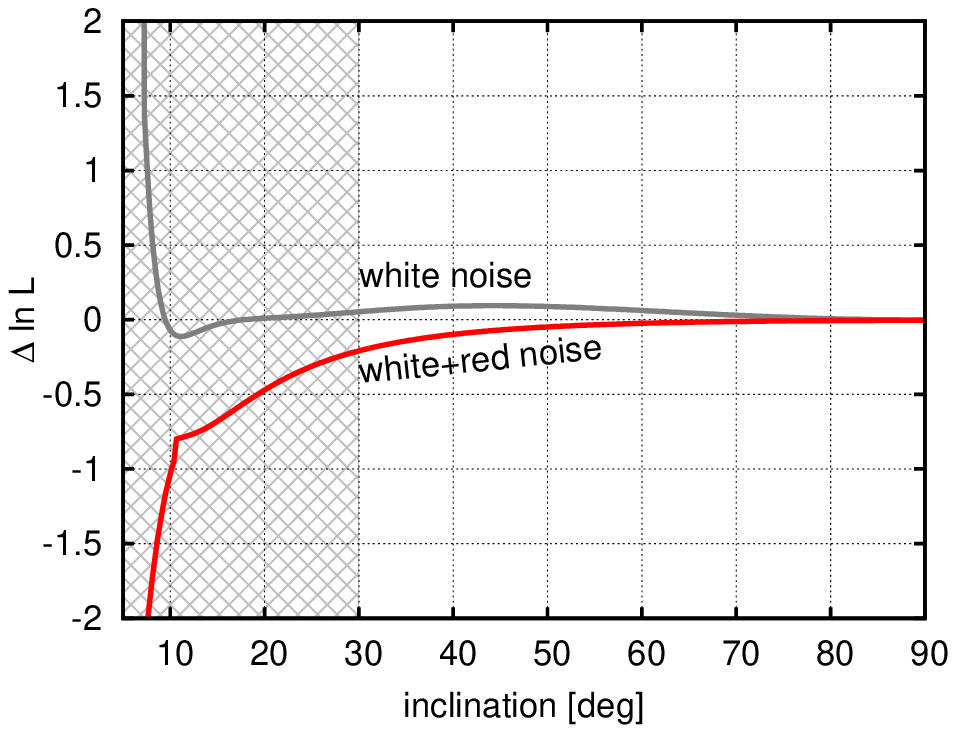}
\caption{The maximized log-likelihood of the 55~Cnc RV data as a function of the
common orbital inclination of the outer planets \emph{b}, \emph{c}, \emph{d}, and \emph{f}.
The innermost planet \emph{e} was not included in the integration. The
log-likelihood curves were shifted vertically so that they both pass through zero at the
abscissa of $90^\circ$. The hashed region labels the instability domain derived from the
results by \citet{Nelson14}, see text.}
\label{fig_55Cnc_inc}
\end{figure}

The orbital inclination in Table~\ref{tab_55Cnc} is fixed at
$90^\circ$ (edge-on) because otherwise it appears undeterminable: we obtain that the value
of the maximized likelihood function changes rather little down to $i$ as small as
$10^\circ$ (see Fig.~\ref{fig_55Cnc_inc}). However, \citep{Nelson14} give the estimation
$i\sim 90^\circ$ with the uncertainty of $\sim 25^\circ$. We view this uncertainty too
optimistic, and we guess that it may owe e.g. to the prior distribution adopted for $i$ by
\citet{Nelson14} in their Bayesian simulation. Note that the
planet 55~Cnc~\emph{e} is transiting its star, so its orbital inclination to the sky plane,
$i_e$, is necessarily close to $90^\circ$. In Fig.~\ref{fig_55Cnc_inc} we vary the
common orbital inclination of the remaining planets $i_{bcdf}$, thus allowing the entire
system to be non-coplanar. According to the results by \citet{Nelson14}, the five-planet
system becomes unstable when the mutual inclination $i_{\rm mut}$ between the orbital plane
of 55~Cnc~\emph{e} and the common orbital plane of the other planets is in the range
from $60^\circ$ to $120^\circ$. As far as $i_e$ is close
to $90^\circ$, the inclination $i_{bcdf}$, considered in Fig.~\ref{fig_55Cnc_inc},
cannot be smaller than $30^\circ$.

Finally, we note that in \citep{Baluev13e} we have done a brief period analysis on the
55~Cnc Lick RV data from \citep{Fischer08}, claiming an uncertain detection of
an additional $9.8$~d signal. The period of $9.8$~d was not confirmed by
\citet{Nelson14} with updated data, and now we do not confirm it too. This $9.8$~d
signal now looks like an artifact of the red noise that was ignored in \citep{Baluev13e}.

\section{Evidences of an activity cycle in Doppler data of 55~Cancri}
\label{sec_activity}
It is remarkable that according to our fit, the ELODIE and HJST datasets possess
considerably different noise correlation timescale $\tau\sim 100$~d, whereas the other
datasets have $\tau\sim 3-15$~d. This may indicate that the correlated noise in the ELODIE
and HJST subsets has essentially different physical nature. Perhaps, it might have
an instrumental origin in ELODIE and HJST, while in the other data it may be caused by the
star activity.

In fact, there is no unique explanation of the differences in the correlation parameters
between RV datasets. First of all, we do not really know how much of the red noise was
generated by the instrumental sources and by the star. But even if the red noise
is entirely caused by the star activity, its numerical parameters may largerly depend on
the characteristics of the spectrograph or telescope, or on the spectrum analysis software.
Also, the differences in the correlation parameters may indicate some temporal variability
of the star activity level.

In the case of 55~Cnc the data are very heterogeneous, coming from different observatories
and teams, so it is difficult to say which effect is responsible for the difference in the
red-noise characteristics. However, if we leave only the Lick and Keck data that
were obtained by means of more or less similar observation technique and spectrum analysis
methods, it appears that the red noise level was close to a minimum in Lick-6 and Keck-1
observations, while in Lick-13, Lick-8 and Keck-2 data it was near a maximum. This
infers an activity cycle of $\sim 13$~yrs, with a single observed minimum near JD2452500
(about 2002.5) and two observed maxima near JD2450000 (about 1996.0) and JD2455000
(about 2009.5).

Various monitoring observations available in public literature on 55~Cnc reveal clear hints
of the long-term activity variation in this star. Results by \citet{Henry00} suggest a
cyclic variation in the Ca II H+K activity indicator with a single observed maximum in
1995. They do not cover the complete cycle with good confidence, but the
activity minimum can be suspected some time after the end of their
observations time-span in 1999. A long-term downtrend was noted by \citet{Marcy02}
in the Str\"omgren photometry between the dates JD2450500 and JD2452500. A later work by
\citet{Fischer08} presents an updated plot of $11$-year photometry series. In this plot we
do not see any long-term flux variation, because it was subtracted by the authors. However,
some long-term cyclic variation in the magnitude of the photometic scatter can be easily
noted. This scatter is likely tied to the spotting activity of the star, which is therefore
variable. If this interpretation is correct, we can see some excessive activity starting
from the beginning of the data and continuing until JD2451400. It is followed by a
depressed activity epoch lasting until JD2453200. After that, an increased activity regime
returns, continuing till the end of the dataset (JD2454200). The apparent minimum is likely
located near JD2452000, and the second maximum is likely near JD2454000.
Unfortunately, \citet{Fischer08} do not give any period estimation or even period analysis
of the long-term flux variability, although they note that ``55 Cnc clearly exhibits
year-to-year variations in mean brightness with an amplitude of 0.001 mag over a timescale
of several years or more''.

These results clearly provide an independent support to our hypothesis that the Doppler red
noise in the 55~Cnc Lick and Keck RV data owes to the star's magnetic activity
cycle, similar to the Solar $11$-year activity variation. Likely, this red noise is caused
by spots evolving on the stellar surface, and the correlation timescale of $\sim 10$~d may
reflect the intrinsic stability of these spots or spot groups. This timescale is also in a
rough agreement with the typical lifetime that we have for sunspots or sunspot groups (days
to weeks). Also, this timescale can be related to star rotation. The rotation period
of 55~Cnc is $\sim 43$~d \citep{Fischer08}, and after $10-20$~d the star would
make $1/4-1/2$ of a single revolution, causing significant changes to the spot pattern seen
on its visible disk, and consequently to the measured radial velocity.

Given this considerations, we undertook an attempt to estimate the period and phase of the
long-term variability of the red noise component in Doppler data. To do this we replace the
noise model~(\ref{rednoise}) with the following modulated modification:
\begin{eqnarray}
V_{ij} = \sigma_{i,\rm wht}^2(p_{\rm wht})\delta_{ij} + V_{ij,\rm red}(p_{\rm red},\tau,P_{\rm m},\lambda_{\rm m}), \nonumber\\
V_{ij,\rm red} = p_{\rm red} A_i(P_{\rm m},\lambda_{\rm m}) A_j(P_{\rm m},\lambda_{\rm m}) R_{ij}(\tau), \nonumber\\
A_i = \cos\left[\frac{\pi}{P_{\rm m}} (t_i-T_0) + \frac{\lambda_{\rm m}}{2} \right], \quad R_{ij} = \rho\left(\frac{t_i-t_j}{\tau}\right).
\label{modrednoise}
\end{eqnarray}
Here $P_{\rm m}$ and $\lambda_{\rm m}$ are the period and the phase of the modulation, and $p_{\rm
red}$ now corresponds to the maximum red-noise magnitude. Note that when $t_i=t_j$ we have
the variance of the red noise component proportional the modulation factor
of $\{1+\cos[2\pi(t-T_0)/P_{\rm m}+\lambda_{\rm m}]\}/2$. This assumes that in the epoch
of minimum activity the red noise vanishes entirely. It is easy to verify that the
matrix $\mathbfss V_{\rm red}$ in~(\ref{modrednoise}) keeps its positive definiteness if it
was positive definite in~(\ref{rednoise}). In~(\ref{modrednoise}) we
assume a regularized model of the white noise, $\sigma_{i,\rm wht}^2$, and require
that $p_{\rm red}>0$.

We include in the analysis only the Lick and Keck data, except for the very old and rather
inaccurate $14$ ``early-era'' Lick data points. The ELODIE and HJST data have too different
value of $\tau$ that may indicate a different source of red noise correlations, possibly
having instrumental nature. Besides, the ELODIE data are rather old and frequently reveal a
systematic annual variations possibly owing to some reduction errors
\citep{Baluev08b,BaluevBeauge14}. Presumably, HET data are located near the epoch of
minimum activity, but in their periodogram we can see a clear power excess at longer
periods (Fig.~\ref{fig_55Cnc_pow}). However, it is mainly generated by a single high
peak at the period of $\sim 200$~d. The time span of these data is very small (the
same $\sim 200$~d), and this variation may be due to some systematic instrumental or
data-reduction drift. Previous studies revealed a systematic annual variation in the HET
data for HD~74156 \citep{Baluev08b,Wittenmyer09,Meschiari11}. Whether the same is true for
the HET data from \citep{Endl12} or not, this RV subset remains too suspicious, and we
belive it is unreliable for this very subtle work.

Thus, we are left with $5$ high-quality RV subsets: Lick13, Lick6, Lick8, Keck1, Keck2. In
the model~(\ref{modrednoise}) we assume that the white-noise term may have different value
of $p_{\rm wht}$ for different RV subsets. The values of $P_{\rm m}$ and $\lambda_{\rm m}$
are assumed the same across the entire RV data, while $p_{\rm red}$ and $\tau$ are assumed
the same for all Lick data and for all Keck data, but are allowed do differ between Lick
and Keck (this looks reasonable based on the data from Table~\ref{tab_55Cnc}). The cross
correlations between different datasets is still set to zero, because otherwise we have to
deal with a non-trivial question, what $\tau$ and $\rho$ we should set for these
cross-correlation matrix blocks of $\mathbfss V$, in particular bearing in
mind the requirement of positive definiteness of $\mathbfss V$. If the red noise is indeed
caused by the star activity then this assumption of zero cross correlations is probably not
very physical, but in practice different Lick and Keck RV subsets either do not overlap or
overlap little, so this assumption does not make significant difference to the final
results. The RV curve is approximated by a five-planet Keplerian model. Bearing in mind the
results presented in Fig.~\ref{fig_55Cnc_rsd}, the effect of the Newtonian planet-planet
interaction is much smaller than the effect of the red noise, and can be therefore
neglected.

\begin{table}
\caption{Best fitting parameters of the 55~Cnc Doppler noise with an activity cycle modulation.}
\label{tab_55Cnc_mod}
\begin{tabular}{@{}ll@{\,}l@{\,}ll@{\,}l@{}}
\hline
                           & Lick 13        & Lick 6         & Lick 8         & Keck 1         & Keck 2         \\
white $\sigma_\star$~[m/s] & $1.72(67)$     & $3.94(59)$     & $3.98(43)$     & $2.79(62)$     & $1.323(80)$    \\
red   $\sigma_\star$~[m/s] & \multicolumn{3}{c}{$9.5(1.1)$}                   & \multicolumn{2}{c}{$3.47(31)$}  \\
$\tau$~[d]                 & \multicolumn{3}{c}{$23.8(7.1)$}                  & \multicolumn{2}{c}{$3.30(85)$}  \\
r.m.s.~[m/s]               & $7.878$        & $5.145$        & $7.035$        & $3.363$        & $3.468$        \\
$P_{\rm m}$~[d]            & \multicolumn{5}{c}{$5610(580)$}    \\
$\lambda_{\rm m}$~[$^\circ$]& \multicolumn{5}{c}{$325(14)$}      \\
\hline
\multicolumn{6}{c}{general characteristics of the fit}   \\
$\tilde l$~[m/s]           & \multicolumn{5}{c}{$3.232$} \\
$d$                        & \multicolumn{5}{c}{$41$}    \\
\hline
\end{tabular}
\end{table}

\begin{figure}
 \includegraphics[width=84mm]{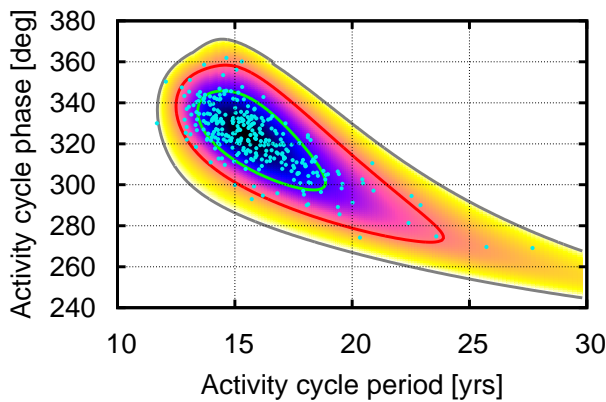}
\caption{Likelihood regions for the period and phase of the 55~Cnc activity cycle, derived
from its Doppler data (see text). The level contours correspond to $1,2,3$-sigma
significance of the likelihood-ratio test. A set of $300$ synthetic best fitting
points generated by Monte Carlo simulation is shown for comparison.}
\label{fig_55Cnc_modsim}
\end{figure}

\begin{figure*}
 \includegraphics[width=0.75\textwidth]{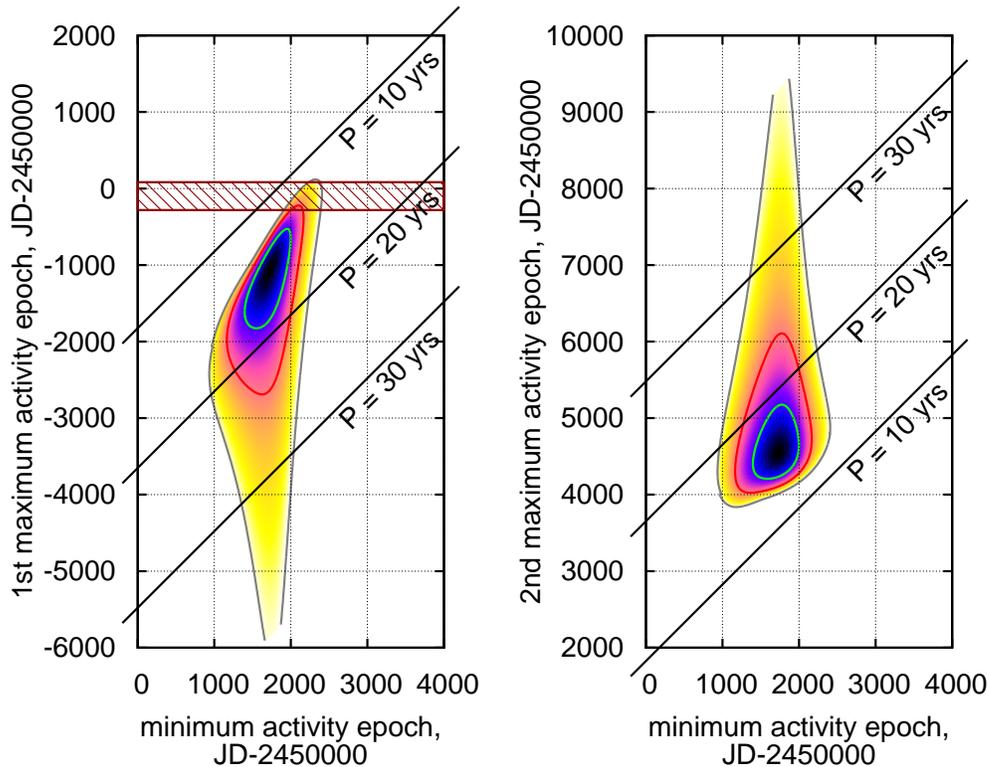}
\caption{Likelihood regions for a single observed minimum and two observed maxima of the
55~Cnc activity, derived from its Doppler data (see text). The hashed chunk in the
left plot labels the year 1995, in which the first maximum was located
by \citet{Henry00} using Ca II H \& K measurements.}
\label{fig_55Cnc_mod}
\end{figure*}

At first we performed the maximum-likelihood fit of the described RV model. We obtained the
best fitting values of $P_{\rm m}=5610\pm 580$ and $\lambda_{\rm m}=325^\circ\pm 14^\circ$
with no remarkable changes in the planetary parameters. We give the other RV noise
estimations in Table~\ref{tab_55Cnc_mod}, omitting the planetary
system parameters. However, it appears that this period estimation is uncertain. The
Doppler data cover the minimum activity epoch of 2001 well, but the two maximum activity
epochs of 1995 and 2009 are covered only partially. Therefore, with only the Doppler
data the uncertainty of $P_{\rm m}$ becomes very asymmetric without a reliable upper limit.
This is demonstrated in Fig.~\ref{fig_55Cnc_mod}, where we show the likelihood regions for
the parameters $(P_{\rm m},\lambda_{\rm m})$. The likelohood contour
levels shown there formally correspond to the asymptotic 1,2,3-sigma significance levels of
the relevant likelihood-ratio test.

For comparison, we also plot over this graph a set of $300$ Monte Carlo points. They
were obtained by assuming that the best fitting model of the real RV data with the implied
noise covarince matrix are true, then simulating in accrodance with this covariance matrix
a set of $300$ synthetic RV datasets, and fitting each of these datasets with the same
model. If the model was significantly non-linear, the simulated points would deviate
from the contours of the likelihood function. However, instead of this we can see a good
agreement. This likely indicated that the model is well linearizable, the fit is
statistically robust, and from the statistical point of view we can safely rely on the
uncertainties derived with the likelihood-ratio test. Unfortunately this does not yet mean
that the best fitting estimations and uncertainties of $P_{\rm m}$ and $\lambda_{\rm m}$
are indeed so reliable in the practical sense. They appear
severely model-dependent and until we understand well the underlying mechanism
of how the red noise is generated in the Doppler data, we should treat these estimations
and confidence domains only as a rough suggestive reference.

In Fig.~\ref{fig_55Cnc_mod} we show similar likelihood regions for two other
equivalent pairs of parameters: (minimum activity epoch + 1st maximum activity
epoch) and (minimum activity epoch + 2nd maximum activity epoch). The activity
minima and maxima are understood here as the minima and maxima
of the Doppler red noise sinusoidal modulation. Anyway, we can see that only the lower
period limit is more or less useful, while the upper one is very vague: even the periods as
large as $30-40$~yrs are allowed.

However, the Ca II H\&K data from \citep{Henry00} yield the epoch of the first
activity maximum with rather good accuracy, and this can be used to dramatically improve
our constraints on $P_{\rm m}$. In the left frame of Fig.~\ref{fig_55Cnc_mod} we
show the horizontal hashed band that refers to the year of 1995, which
is the maximum activity year from \citep{Henry00}. We refitted our model with a constraint
that the phase of the modulation is zero at JD2449800 (some date in 1995). This suggested a
refined estimation $P_{\rm m} = 12.6 \pm^{2.5}_{1.0}$~yrs (uncertainties are
from the likelihood-ratio test). Thus, the 55~Cnc activity cycle is similar to the Solar
cycle or slightly longer. The next year of 2015 is a predicted year of minimum activity of
55~Cnc, although we emphasize that this prediction, as well as the period estimation above,
are sensitive to model assumptions, in particular to the assumption that the activity is
strictly periodic.

In Table~\ref{tab_55Cnc_mod} we can see that the Doppler red noise in the Lick and
Keck data have considerably different parameters: both the correlation timescale $\tau$ and
the maximum red-noise variance $\sigma_{\star,\rm red}$ are significantly larger for the
Lick data. The difference in $\tau$ looks rather
natural, because the Lick/Hamilton spectrograph likely can ``see''
only the bigger, and thus more long-living, spots and spot groups on the star disk due
to the weaker techinical characteristics. However, the difference in the red noise
magnitude is rather paradoxical. It is unexpected that a weaker instrument is more
sensitive to a subtle spot activity. We have several explanations to this:
\begin{enumerate}
\item Lick data contain an additional red-noise component that may have instrumental nature.
\item The star was more active in the maximum of 1995 than in that of 2009.
\item Some hidden technicalities can suppress the red noise in the Keck data. This may
be broadly analogous to the long-exposure averaging effect on the
short-period astroseismology harmonics.
\item The maximum-likelihood procedure of obtaining the Doppler RV value
from the raw spectrum used in \citep{Fischer14,Nelson14}
is clearly non-linear, and non-linear algorithms are known to generate negative feedbacks.
\end{enumerate}

The estimated activity period is suspiciously close to the period of the outermost and most
massive planet~\emph{d}. Can this modulated red RV noise be caused by e.g. Newtonian
perturbations from this planet on the motion of the inner planets? According to the results
of Fig.~\ref{fig_55Cnc_rsd}, the effect of the
correlated noise exceeds the Newtonian perturbations for $i=90^\circ$ by a factor of $\sim
5$. Thus these two effects may become comparable only for $i$ as small
as $10^\circ$, scaling the real planetary masses up enough. However, even in this case
there are several arguments against such interpretation of the red RV noise in the data:
\begin{enumerate}
\item The dynamics of the system on so short time span is definitely regular, implying that
the frequency spectrum of the Newtonian perturbations in the radial velocity should contain
only a few isolated frequencies tied to the planetary orbital mean motions and their linear
combinations, or to the secular periods. The red noise that we can see in periodograms
of Fig.~\ref{fig_55Cnc_pow} does not demonstrate this.

\item The magnitude of perturbations and other their characteristics should be the
same for the Lick and Keck data, but this does not holds for the red noise that we see.

\item The inclination of $\sim 10^\circ$ is inconsistent with the results by \citet{McArthur04},
who give for the outermost planet $i=53^\circ$ based on HST astrometry + RV data available
that time. The magnitude of the astrometric displacement should scale roughly as
$\propto 1/\sin i$ (due to an increase of the real planetary masses), meaning that
if $i$ was only $10^\circ$ then \citet{McArthur04} would be wrong by a factor of $\sim 4$,
or even larger due to the projection effect. This does not seem likely, even if the results
by \citet{McArthur04} were based on an old and inaccurate value of $P_d$.

\item There is a clear correlation of the derived Doppler activity cycle with the Ca
II H\&K and photometry observations discussed above.
\end{enumerate}
Therefore, we treat the coincidence between the activity cycle and the period of the
outermost planet as accidental. Interestingly, the Solar activity cycle is close
to the Jupiter orbital period too, and in the 55~Cnc system the period
of the planet~\emph{b} is close to the star rotation period. From the other side, some
discussion is available in the literature investigating the relationship between the Solar
activity and planetary orbital motion \citep{Abreu12,Cameron13,TanCheng13}. Our estimation
of the 55~Cnc activity cycle may offer some fresh source material to a work of such type.

\section{Conclusions}
The radial-velocity method of exoplanets detection has encountered a natural limit set
by the intrinsic activity of even relatively quiet Solar-type stars. Recent studies reveal
that careful modelling of the Doppler noise may be a key to the future progress of the
radial-velocity exoplanet detections. For example, \citet{Dumusque12} were able to
reveal the planet orbiting $\alpha$~Cen~B only after suppressing the star's rotation and
long-term activity effects in their Doppler data (although this detection currently appears
model-dependent, see \citet{Hatzes13}). In \citep{Baluev13a} the
planet GJ~581~\emph{e} could be confirmed in the Keck data only after taking the red noise
into account. From the other side, an application of advanced noise models may lead to
planet retractions too: see the cases of GJ~581~\emph{f} and~\emph{g} in
\citep{Baluev13a,Tuomi12}, and of GJ~581~\emph{d} in \citep{Robertson14}, and likely the
case of the GJ~667~C multiplanet system in \citep{FerozHobson14}.

The present paper further emphasizes the importance of a wise selection of the Doppler
noise model and the need to increase the complexity of the latter. For example, in
Table~\ref{tab_55Cnc}, about 2/3 of the model parameters is related to the Doppler
noise, while only 1/3 of them refers to the exoplanetary system. Moreove, this work reveals
that it is not enough to think only of the formal accuracy and physical justification of
the RV noise model. Additionally to this, we must take care of mathematical smoothness
of the resulting likelihood function, and carefully define the admissible parametric
domain, paying a special attention to the regions where the noise model is
formally not physical but still mathematically defined.

The 55~Cnc case appears even more plentiful in what concerns the usefulness of non-standard
noise models. Other solar-type stars exist, in which an activity cycle was detected by
means of the Doppler RV monitoring \citep{Dumusque11}, but 55~Cancri is likely among the
most detailedly investigated ones. Probably, more practical methods stellar activity
monitoring are the the Ca II H\&K lines measurements \citep{Henry00,Robertson13} and
high-accuracy photometry \citep{Fischer08}. Since these methods allow more
easy and accurate way to track the stellar activity, they may be useful in construction of
a more reliable model of the red Doppler noise. Instead of performing a pure Doppler fit
that would rely on some particual temporal model of the long-term modulation, we may try to
subtract this modulated red noise based on its correlation with less expensive
observations. This would be generally similar to what \citet{Dumusque12,Tuomi14}
suggest, but the activity index is correlated with the magnitude of the red RV noise
instead of the raw radial velocity. Thus it is still necessary to answer the question,
which approach is more appropriate in practice. In any case, a good reduction of the
stellar activity hints in Doppler data would likely increase the
accuracy and reliability of the derived exoplanetary model for the stars like 55~Cnc.

An RV fitting algorithm based on our regularized jitter model is now implemented
in the PlanetPack package \citep{Baluev13c} and will be available in the forthcoming
version PlanetPack~2.0. This version will also include two other noise
models discussed
above: the multiplicative~(\ref{mult}) and the truncated additive~(\ref{addtrunc}) one. The
pure additive model~(\ref{add}) can be obtained from the regularized or truncated additive
ones by setting the regularization or truncation parameter
to zero. The regularized white-noise model is now the default choice in PlanetPack, because
so far we noticed no bad side or trade-off consequences from its usage: the fit robustness
is either improved or left unchanged. In view of that, the additive model may now be
considered as deprecated. The model of the red noise in PlanetPack is now corrected to
disallow negative values of the parameter $p_{\rm red}$ in~(\ref{rednoise}). PlanetPack 2.0
can also deal with the modulated red noise~(\ref{modrednoise}), although this is still
an experimental feature. To achieve the best performance on modern multi-core CPUs, some
computationally-heavy PlanetPack algorithms can now be parallelized in multiple threads.

In addition to the new features that are based on the methods from the present
paper, PlanetPack 2.0 will include a new optimized algorithm of evaluating the Keplerian
periodogram of the RV data \citep{Cumming04}, equipping it with a new fast analytic method
of estimating its significance levels \citep{Baluev14b}. Finally, it
includes a maximum-likelihood algorithm for fitting exoplanetary transit lightcurves using
the new accurate limb-darkening model by \citet{AbubGost13}. Although still rather
experimental, this transit fitting module inherits most noise modelling features of the RV
fitter, including an option of handling the red noise in the photometry data
with the model~(\ref{rednoise}) and the use of any of the three white noise models
mentioned above.

\section*{Acknowledgements}
This work was supported by the Russian Foundation for Basic Research (projects No.
12-02-31119 mol\_a and 14-02-92615 KO\_a), by the President of
Russia grant for young scientists (MK-733.2014.2) and by the programme of the Presidium of
Russian Academy of Sciences ``Non-stationary phenomena in the objects of the Universe''. I
would like to express my gratitude to the reviewer, G.~Anglada-Escud\'e, for his invaluable
suggestions concerning the draft of the paper.

\bibliographystyle{mn2e}
\bibliography{regula}

\appendix

\section{A weighted analogue of the Kolmogorov-Smirnov test for tail probabilities}
\label{sec_wKS}
Consider a sequence of $N$ random numbers $x_i$ sampled from some parent
distribution.\footnote{Within this appendix these designations are unrelated to the number
of observations and measurements in a time series.} Our task is to test whether this
observed set of $x_i$ is consistent with some proposed cumulative distribution function
(hereafter CDF) $F(x)$ or not. This can be verified by means of the classic
Kolmogorov-Smirnov (hereafter K-S) test based on the maximum deviation between the empiric
CDF $F^*(x)$, which is constructed from $x_i$, and the proposed CDF $F(x)$:
\begin{equation}
\varkappa = \sqrt N \max |F^*(x)-F(x)|.
\label{KSstat}
\end{equation}
Under assumption that $F(x)$ is continuous and indeed represents the CDF of $x$, and for
$N\to\infty$, the distribution function $P(\varkappa)$ of the K-S test statistic
$\varkappa$ is given by the following decomposition, first obtained by A.N.~Kolmogorov:
\begin{equation}
P(\varkappa) = 1 + 2 \sum_{n=1}^{\infty} (-1)^n e^{-2 n^2 \varkappa^2}.
\label{KSdistr}
\end{equation}
For a given confidence probability $p$, the quantiles $\varkappa^*(p)$ of the latter
distribution may be used to construct the $p$-value CDF band as
\begin{equation}
\left[ F(x)-\frac{\varkappa^*}{\sqrt N}, \quad F(x)+\frac{\varkappa^*}{\sqrt N} \right].
\end{equation}
If the empiric CDF $F^*(x)$ always lies within this band then our sample is consistent with
the selected $F(x)$, and otherwise we conclude the $F(x)$ does not describe the observed
sample well. Note that the quantiles $\varkappa^*$, corresponding to the one-, two-, and
three-sigma significance, are equal to $0.96$, $1.38$, and $1.82$.

It is important for the validity of the test that $F(x)$ should be set \emph{a priori} as
a plain function of only $x$: it is illegal to first adopt some \emph{parameterized} model
$F(x,\theta)$ with unknown $\theta$, fit it via $\theta$ on the basis of $x_i$, and then
apply the K-S test to the same sample $x_i$.

The strength of the K-S test is in its simplicity and in the invariance of~(\ref{KSdistr})
with respect to $F$. However the K-S test is known for its low power: as a trade-off to its
simplicity and generality, it has rather poor sensitivity to small differences between $F$
and $F^*$. The latter sensitivity quickly degrades in the tails of the distribution,
i.e. when $F(x)$ is close to either zero or unit.

Let us demonstrate this. Considering that $x$ is fixed, the values of $NF^*(x)$ obey the
binomial distribution with the probability parameter equal to $F(x)$. It is easy to show
that for large $N$ this binomial distribution can be approximated by a Gaussian one with
\begin{equation}
\expect F^*(x) = F(x), \quad \disp F^*(x) = F(x)(1-F(x))/N.
\end{equation}
Clearly, the natural uncertainty of $F^*(x)$ is not uniform in $x$, decreasing to the
tails of the distribution. However, the K-S statistic~(\ref{KSstat}) does not take this
behaviour into accout, as if assuming that the uncertainty of $F^*(x)$ was constant.

When dealing with simulated $\FAP$ curves, we are especially interested in the tails of the
distribution, corresponding to small $\FAP$s. In this domain the K-S test generates hugely
overestimated confidence ranges, making us essentially blind to small but in principle
detectable deviations.

This problem was considered by \citet{ChichBouch12}, who suggest a weighted modification
of the K-S statistic~(\ref{KSstat}):
\begin{equation}
\tilde\varkappa = \sqrt N \max_{x: a\leq F(x)\leq b} \frac{|F^*(x)-F(x)|}{\sqrt{F(x)(1-F(x))}}.
\label{modKSstat}
\end{equation}
According to \citet{ChichBouch12}, the asymptotic distrubution function of
$\tilde\varkappa$ is different from~(\ref{KSdistr}), and obtaining it appears a pretty
hard task, involving the theory of random processes and stochastic differential equations.
The final result looks like:
\begin{equation}
P(\tilde\varkappa) \simeq A(\tilde\varkappa) e^{ - T \theta(\tilde\varkappa)},
\label{modKSdistr}
\end{equation}
where
\begin{equation}
T = \ln \sqrt{\frac{b(1-a)}{a(1-b)}}
\end{equation}
depends on the desired probability limits $a$ and $b$ and is assumed large, and the
functions $A$ and $\theta$ obey the asymptotic behaviour
\begin{eqnarray}
A(k) &\simeq& {\mathop{\rm erf}\nolimits}^2\left(\frac{k}{\sqrt{2}}\right), \nonumber\\
\theta(k) &\simeq& \sqrt{\frac{2}{\pi}} k e^{-k^2/2}
\end{eqnarray}
for large argument. Note that the requirement for $T$ and $k$ to be ``large'' is
understood in a pretty mild sense here.

Given some quantile $\tilde\varkappa^*(p)$, corresponding to the confidence probability
$p$, we may construct the associated confidence band
\begin{equation}
\left[ F - \tilde\varkappa^* \sqrt{\frac{F(1-F)}{N}},\quad F + \tilde\varkappa^* \sqrt{\frac{F(1-F)}{N}}\, \right].
\end{equation}
If our $F^*(x)$ stays inside this band everywhere within the abscissa range corresponding
to $a \leq F(x)\leq b$, the observed sample is consistent with $F(x)$. Otherwise, it is
not. In Sect.~\ref{sec_simul} we mainly deal with $\FAP$s, which are complementary
probabilities to $F$, and with $\FAP$ ratios. The confidence range for the $\FAP$ ratio
$(1-F^*)/(1-F)$ is given by
\begin{equation}
\left[ 1 - \tilde\varkappa^* \sqrt{\frac{F}{N(1-F)}},\quad 1 + \tilde\varkappa^* \sqrt{\frac{F}{N(1-F)}}\, \right].
\end{equation}

\citet{ChichBouch12} eventually apply their results to the case with $a=1/N$ and
$b=(N-1)/N$, corresponding to the absolute minimum and maximum values of $F^*$. We however
believe that so wide limits might be dangerous: the Gaussian approximation for $F^*$ is
invalid so deep in the tails of the distribution, where $F^*$ actually behaves as a
Poisson variate. Besides, we have no practical need to move so far in the tail, since
these values of $F^*$ are unreliable in any case. The $\FAP$ graphs plotted in
Sect.~\ref{sec_simul} above are truncated at the abscissa of $\FAP=10^{-3}$, implying the
upper probability limit $b=0.999$. Note that since we have $N=60000$, this probability
corresponds to $N\, \FAP=60$ false alarm events, which is large enough to justify the
Gaussian approximation of $F^*$. For the lower limit we adopt $a=0.9$, since we have no
practical need to consider $\FAP>0.1$ (it is expected and unsurprising that our analytic
$\FAP$ approximations are significantly overestimated there). These limits $a$ and $b$
correspond to $T=2.35$. Substituting this value of $T$ in the formulae above, we obtain
(numerically) that the critical values of $\tilde\varkappa$ that map to the one-, two-,
and three-sigma confidence levels, are equal to $2.26$, $3.14$, and $4.00$. In the plots
of Sect.~\ref{sec_simul} we show the confidence bands corresponding to these critical
values.

\bsp

\label{lastpage}

\end{document}